\documentclass[aoas]{imsart}

\RequirePackage{amsthm,amsmath,amsfonts,amssymb}
\RequirePackage{natbib}
\RequirePackage[colorlinks,citecolor=blue,urlcolor=blue]{hyperref}
\RequirePackage{graphicx}

\usepackage{amsmath, amssymb}
\usepackage{bbold}
\usepackage{mathbbol}
\usepackage{algorithm}
\usepackage{algorithmic}
\usepackage{enumitem}
\usepackage{float}
\usepackage{xcolor}
\usepackage{bm}
\usepackage{multirow}
\usepackage{lscape}
\usepackage{rotating}
\usepackage{adjustbox}
\usepackage{booktabs}
\usepackage{ragged2e}
\newtheorem{theorem}{Theorem}
\newtheorem{proposition}{Proposition}
\usepackage{caption}
\usepackage{subcaption}
\usepackage{float}
\def\*#1{\mathbf{#1}}
\def\+#1{\amsmathbb{#1}}
\def\^#1{\mathbb{#1}}
\newcommand{\tno}{\mbox{TNO$_3$}}

\startlocaldefs
\endlocaldefs

\begin{document}

\begin{frontmatter}
\title{Spatially-Indexed Longitudinal Distributional Outcome Regression for Environmental Monitoring}
\runtitle{Spatially-Indexed Longitudinal DOR}

\begin{aug}
%%%%%%%%%%%%%%%%%%%%%%%%%%%%%%%%%%%%%%%%%%%%%%%
%% Only one address is permitted per author. %%
%% Only division, organization and e-mail is %%
%% included in the address.                  %%
%%%%%%%%%%%%%%%%%%%%%%%%%%%%%%%%%%%%%%%%%%%%%%%

\author[A]{\fnms{Rahul}~\snm{Ghosal}\thanksref{t1}\ead[label=e1]{rghosal@mailbox.sc.edu}}
\author[B]{\fnms{Suman}~\snm{Majumder}\thanksref{t1}\ead[label=e2]{smajumder@isical.ac.in}}
%\and
\author[C]{\fnms{Indranil}~\snm{Sahoo}\ead[label=e3]{sahooi@vcu.edu}}

\thankstext{t1}{Co-first authors.}

%%%%%%%%%%%%%%%%%%%%%%%%%%%%%%%%%%%%%%%%%%%%%%
%% Addresses                                %%
%%%%%%%%%%%%%%%%%%%%%%%%%%%%%%%%%%%%%%%%%%%%%%
\address[A]{Department of Epidemiology and Biostatistics, University of South Carolina, Columbia, SC, USA\printead[presep={,\ }]{e1}}

\address[B]{Interdisciplinary Statistical Research Unit, Indian Statistical Institute, Kolkata, WB, India\printead[presep={,\ }]{e2}}

\address[C]{Department of Mathematics and Statistics, Virginia Commonwealth University, Richmond, VA, USA\printead[presep={,\ }]{e3}}
\end{aug}

\begin{abstract}
Characterizing longitudinal changes in region-specific distribution of environmental exposures, such as total nitrate ($\tno$) concentrations, is critical for understanding localized ecological risks that are otherwise obscured by standard mean-level modeling. However, modeling longitudinal distributional outcomes across spatial regions presents significant methodological challenges. The random objects are spatio-temporally dependent, and mathematical constraints are inherent to distributional representations, such as, monotonicity of quantile functions. To address this, we propose a novel spatially-indexed longitudinal distributional outcome regression model. The distributional coefficients corresponding to the fixed effects of covariates are modeled using Bernstein basis polynomials, while spatio-temporal random effects are flexibly captured via tensor product expansions of splines. We develop a scalable Markov Chain Monte Carlo (MCMC) algorithm to explicitly account for spatial dependencies, and introduce a fast two-stage projected-posterior approach to preserve the monotonicity of the predicted subject-specific quantile functions. Extensive simulation studies demonstrate that the proposed framework achieves superior estimation accuracy and predictive performance compared to standard non-spatial distributional outcome regression. We apply our methodology to predict monthly, site-specific distributions of $\tno$ concentrations across the contiguous United States. Accounting for spatial correlation provides substantially lower uncertainty in the estimated distributional effects and improves predictive performance over the non-spatial alternative, offering a robust, interpretable tool for spatio-temporal environmental monitoring.
\end{abstract}


\end{frontmatter}

%%%%%%%%%%%%%%%%%%%%%%%%%%%%%%%%%%%%%%%%%%%%%%
%%%% Main text entry area:
%%%%%%%%%%%%%%%%%%%%%%%%%%%%%%%%%%%%%%%%%%%%%%

\section{Introduction}
The advent and widespread deployment of continuous, high-frequency sensor technology has led to a paradigm shift in environmental and ecological monitoring over the past few decades. These advanced observational networks systematically enable tracking of a diverse array of environmental exposures, e.g., fine particulate matter (PM$_{2.5}$), ozone, sulfates, and ammonium, alongside high-resolution meteorological variables across broad geographic domains. Prominent national network programs, such as the Clean Air Status and Trends Network (CASTNet) \citep{epa_castnet} maintained by the United States Environmental Protection Agency (EPA), continuously produce massive repositories of spatio-temporal data \citep{baumgardner1998rural,ghosh2010spatio,bondell2010joint}. The increasing availability of such high-frequency data is fundamentally transforming ecological risk assessment, enabling researchers to study highly localized volatility in air quality and environmental exposures.

Traditional statistical approaches and deterministic numerical air quality models applied to these massive environmental datasets predominantly focus on modeling the conditional mean or specific scalar summaries (maximum or minimum) of the pollutant concentrations, aggregated at the daily, weekly, or monthly levels based on the research question of interest. However, characterizing the full distribution of such environmental exposures \citep{dhond2022advancing} can be of paramount importance for comprehensive environmental monitoring and ecological risk assessment. While mean-level modeling is effective for identifying general trends and interpretability, severe environmental and ecological hazards are rarely driven or captured by average conditions. Acute respiratory health risks, or severe water quality degradation are often triggered by extreme exposure events, which are captured exclusively in the upper tails (e.g., the 90th or 95th percentiles) of the pollutant distribution. Furthermore, each site or location can have its own site-specific exposure distribution, which often evolves longitudinally over months and seasons, exhibiting complex, non-stationary temporal dynamics. To properly characterize and mitigate localized ecological risks, we propose to shift the analytical focus from scalar summaries to modeling the complete, site-specific distributions of these exposures as they evolve longitudinally over time.

Among the various pollutants tracked by environmental networks, total nitrate ($\tno$) constitutes a major fraction of fine particulate matter (PM$_{2.5}$) across the United States \citep{malm2004spatial,ghosh2010spatio}. However, despite its prevalence and severe environmental impact, $\tno$ remains one of the most difficult atmospheric components to accurately simulate using deterministic numerical air quality models \citep{yu2005assessment,appel2008evaluation,ghosh2010spatio}. Critically, the difficulty of simulating $\tno$ is not merely one of level, but of mechanism. Distinct formation and loss pathways operate on different parts of the concentration distribution and at different times of year. Because these pathways may reshape the distribution rather than simply translate it, characterizing how meteorological surrogates act differentially across the entire site-specific distribution, and not just its mean or a single quantile, is essential for attributing variability to physically interpretable processes. Accurately capturing these empirical relationships requires statistical models capable of accommodating the spatial and temporal heterogeneity observed in such data. By treating the high-frequency site-specific records of $\tno$ as complete distributions rather than simple scalar averages, researchers can investigate how various meteorological covariates exhibit heterogeneous association with the extremes versus the median, or any other part of the nitrate distribution over time within a unified modeling framework.

\subsection{Motivating Data Application} \label{sec:motivation}

Our method is motivated by a large-scale environmental dataset tracking total nitrate ($\tno$) concentrations across the contiguous United States from $2010$ to $2019$. Using continuously monitored data from 92 monitoring stations of the Clean Air Status and Trends Network (CASTNet), we are interested in modeling region-specific monthly  $\tno$ distributions and understanding the heterogeneous effects of meteorological covariates across different parts of the response distribution.

\begin{figure}[ht]
    \centering
    
    % Panel (a): Median
    \begin{subfigure}[b]{0.7\textwidth}
        \centering
        \includegraphics[width=1\linewidth]{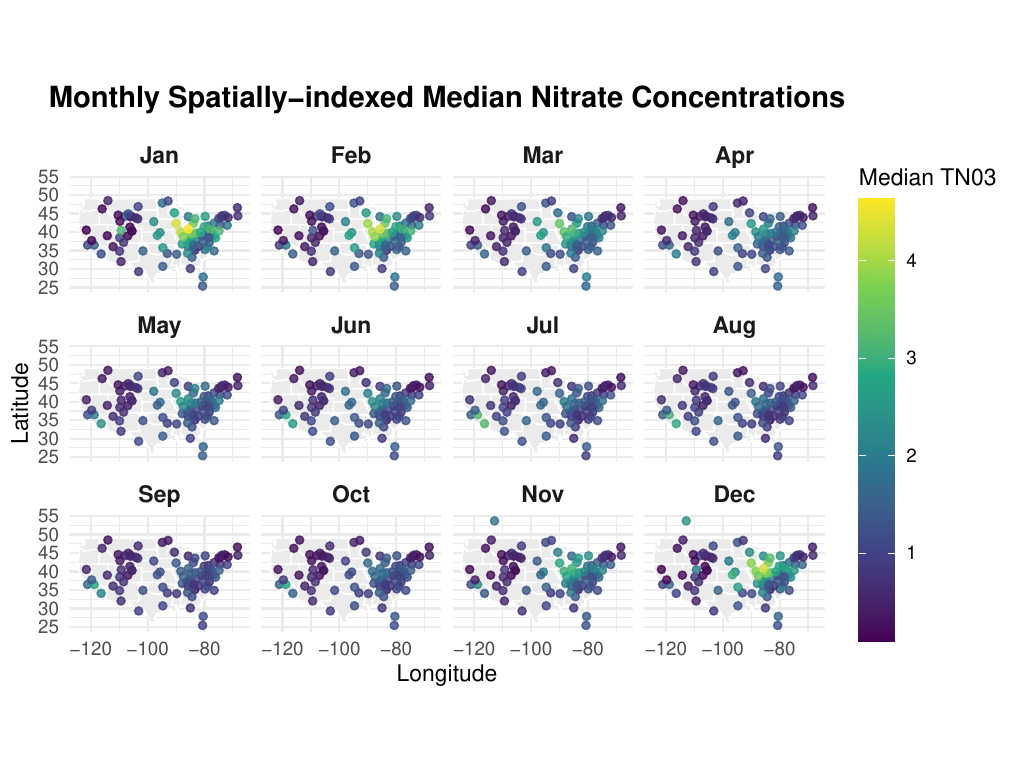}
        %\caption{Median total nitrate ($\tno$) concentration.}
        \label{fig:tn03_median}
    \end{subfigure} 
    
    \vspace{-1 cm}
    
    \begin{subfigure}[b]{0.7\textwidth}
        \centering
        % Replace with your actual filename for the 95th percentile plot
        \includegraphics[width=1\linewidth]{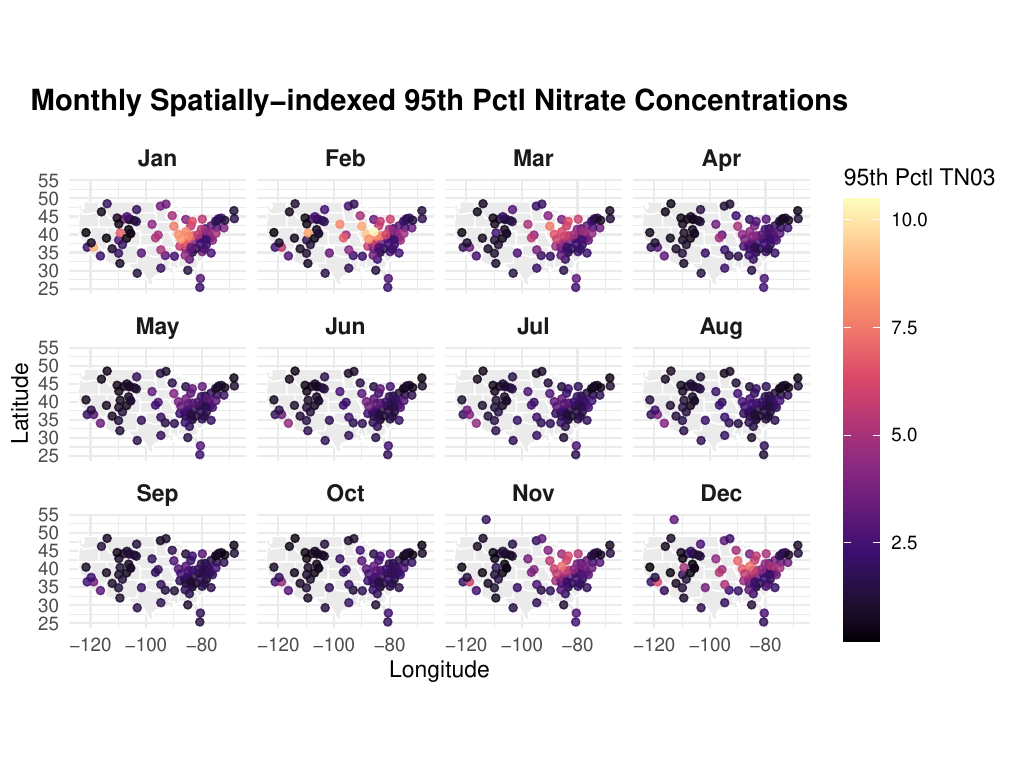} 
        %\caption{95th percentile total nitrate ($\tno$) concentration.}
        \label{fig:tn03_q95}
    \end{subfigure}
    
    \caption{Monthly spatially-indexed distribution of total nitrate ($\tno$) concentrations across the United States based on data from 2010-2019. The top panel illustrates the median concentrations. The bottom panel displays the  95th percentile of the concentration.}
    \label{fig:tn03_spatial_heterogeneity}
\end{figure}

Figure \ref{fig:tn03_spatial_heterogeneity} displays the spatially varying site-specific median (top panel) and 95th percentiles of $\tno$ concentrations (bottom panel) across 12 calendar months.  It is clear from the data that both the median and the 95th percentile exhibit pronounced spatial structure, with a high-concentration cluster in the Midwest and along the mid-Atlantic corridor, while stations in the Pacific Northwest and Rocky Mountain regions are among the lowest concentrations. There is substantial monthly (seasonal) variation superimposed on this spatial pattern, as both median and 95th percentiles of nitrate concentrations are elevated in late winter and spring, and decline through the summer and fall. %It is expected that local climate covariates, e.g, average monthly temperature and total precipitation, would impact these two parts of the nitrate distribution differently. 
Although this spatio-seasonal pattern is attributable to the underlying chemistry driven by formation and loss pathways such as gas/particle partitioning and photochemical production, these pathways are themselves modulated by local climate. 
%Temperature serves as a surrogate for the partitioning--photochemistry balance, with cold conditions favoring particulate accumulation in the upper tail and warm conditions favoring central-mass photochemical production, while precipitation acts as a wet-removal scavenger expected to depress concentrations across the distribution. Because these competing influences act differentially across the $\tno$ distribution rather than uniformly shifting it, understanding how temperature and precipitation independently shape the median and the tails of the distribution, while appropriately borrowing strength across spatial neighbors, is crucial for assessing ecological vulnerability.
This motivates a unified framework that models the region- and month-specific $\tno$ distributions as a function of the climate covariates, provides interpretable effects on the entire outcome distribution, and supports prediction of the monthly $\tno$ distribution in unmonitored locations.

%These competing influences are expected to act differentially on the lower versus upper tails of the $\tno$ distribution}. Understanding how temperature and precipitation independently affect the lower versus upper tails of the $\tno$ distribution, while appropriately borrowing strength across spatial neighbors, is crucial for assessing ecological vulnerability. Instead of modelling these outcomes separately, it is of interest to come up with a unified framework that models the region and month-specific $\tno$ distributions as a function of the climate covariates, provides interpretable effects on the entire outcome distribution, and can be useful for predicting the monthly $\tno$ distribution in unmonitored locations.

\subsection{Related Work} \label{sec: lit_review}

With the increasing availability of high-frequency environmental sensor networks, it is now possible to treat site-specific monthly records of pollutant concentrations as distributional observations \citep{petersen2021modeling,ghosal2023distributional} rather than scalar summaries. Distributional Data Analysis (DDA) is an emerging field in Statistics \citep{petersen2016functional} %,petersen2019frechet,yang2020quantile,yang2020random,matabuena2020glucodensities,matabuena2023distributional,ghosal2023distributional,ghosal2025distributional} 
that provides a rigorous framework for analyzing such intensive repeated measures data, encoding the underlying distribution through histograms, densities, quantile functions, and various such distributional representations.%, yang2020random,ghosal2023distributional,ghosal2025distributional}.
DDA has seen successful applications across wearable digital health technologies, including physical activity and actigraphy \citep{Ghosal2022, matabuena2023distributional, matabuena2025predicting}, continuously monitored glucose \citep{matabuena2020glucodensities,  katta2024interpretable, coulter2025fast,matabuena2026multilevel}, and heart rate \citep{ghosal2025distributional}, demonstrating that distributional representations yield richer scientific insights than scalar summaries. In our motivating application, we are interested in modeling the site and month-specific $\tno$ distribution as an outcome, which carries  meaningful information beyond the mean or aggregated summary metrics. Despite this natural fit, the development of DDA methods tailored to spatio-temporal environmental data remains limited.

Several regression models have been developed with distributional outcomes and scalar or distributional predictors. \citet{yang2020quantile} proposed a quantile-function-on-scalar regression model that represents distributional outcomes via subject-specific quantile functions and models them as functional responses as a linear combination of scalar predictors. \citet{yang2020random} extended this by enforcing monotonicity of the quantile functions via I-splines \citep{ramsay1988monotone} applied to individual regression coefficients. \citet{ghosal2025distributional} developed a distributional outcome regression via quantile functions framework that achieves jointly monotone predicted quantile functions under minimal shape constraints on regression coefficients through Bernstein polynomial expansions, and provided asymptotic projection-based joint confidence bands. Fr\'{e}chet regression \citep{petersen2019frechet} provides a nonparametric approach for distributional outcomes in metric spaces, and has been extended to accommodate causal inference \citep{lin2023causal, katta2024interpretable}, variable selection \citep{coulter2025fast}, single-index models \citep{bhattacharjee2023single}, and multilevel, time-varying and longitudinal designs \citep{dubey2020functional,bhattacharjee2025geodesic,matabuena2026multilevel}. Distribution-on-distribution regression has also been developed via optimal transport \citep{ghodrati2022distribution} and Wasserstein regression-based \citep{chen2021wasserstein} approaches. All of these methods assume independence across the subjects or the fundamental units of observations, which is not feasible in environmental monitoring settings where spatial proximity and the longitudinal nature of the observations induce dependence among the unit of measurements.

In the spatial statistics literature, models for spatially indexed functional data have received
considerable attention \citep{delicado2010statistics, giraldo2010continuous,
nerini2010cokriging}. Spatio-temporal models for environmental data have a rich history focused on scalar or multivariate Gaussian outcomes \citep{cressie2011statistics, banerjee2014hierarchical} and have been extended to functional responses \citep{baladandayuthapani2008bayesian}. Functional Kriging and spatially varying coefficient models have been developed for spatial interpolation and prediction of functional outcomes \citep{giraldo2010continuous}. Functional CAR models have also been developed for large spatially correlated datasets via a basis-transformation approach \citep{zhang2016functional}. Recent works have modeled multilevel functional data with spatial dependence, including fast methods for spatially correlated multilevel functional data \citep{staicu2010fast}, multilevel Bayesian spatio-temporal functional models that induce spatial dependence through a Karhunen-Lo\`{e}ve expansion with a conditional autoregressive (CAR) structure on region-specific principal component scores \citep{li2021multilevel} and its varying-coefficient extension \citep{li2022multilevel}. These methods model functional responses with spatial dependence, but none targets monotone distribution-valued outcomes.

Another related line of work models the conditional population quantiles of a scalar environmental response across space, including the Bayesian spatial \citep{reich2011bayesian} and spatio-temporal \citep{reich2012spatiotemporal} quantile regression. This framework uses a Bernstein polynomial basis expansion to guarantee non-crossing of quantile estimates. However, we emphasize here that our proposed framework of distributional regression is different from quantile regression. Rather than modeling the population quantiles of the outcome, we are modeling the site-specific distribution of a continuously monitored outcome, for which we adopt a quantile function representation. Finally, \citet{ghosh2010spatio} and \citet{bondell2010joint} both modeled the CASTNet $\tno$ data as a scalar log-transformed response, the former employing reparameterized dynamic space-time models with temporally evolving regression coefficients and the latter developing joint fixed and random effect selection within a linear mixed model. Neither of these methods is designed to handle distributional (e.g., site-specific monotone quantile function valued) outcomes. The extension to spatially indexed and longitudinally varying distributional outcomes presents significant statistical and computational challenge, since spatio-temporal random effects must be incorporated in a manner that respects the non-decreasing constraint on predicted quantile functions, and must guarantee the required shape constraints \citep{ghosal2025distributional}.

\subsection{Our contributions}
To the best of our knowledge, no existing statistical framework models spatially-indexed longitudinally varying distributional outcomes with flexible and interpretable covariate effects as needed for our motivating application. To fill this gap, we propose a spatially-indexed longitudinal distributional outcome regression (SILDOR) model that advances the literature in three key respects. First, we develop a Bayesian regression framework for distributional outcomes that are spatially indexed and longitudinal in nature. To our knowledge, this is the first regression model for spatially-indexed, longitudinal distribution-valued outcomes which addresses both spatio-temporal dependence and distributional constraints simultaneously. Second, we introduce a spatio-temporal random effects structure, in which site-specific and longitudinal distributional deviations from the global fixed effects are modeled via a tensor product of Bernstein basis functions (in the quantile direction) and spline basis functions (in the time direction), with spatially correlated coefficients governed by a Mat\'{e}rn covariance \citep{matern1960spatial} (Section \ref{sec:methods}). This structure enables nonparametric, data-adaptive borrowing of strength across neighboring locations while capturing temporal dynamics. Third, and critically, we develop a fast two-stage projected-posterior algorithm for Bayesian inference that avoids the computational burden of a directly constrained Markov Chain Monte Carlo (MCMC) while guaranteeing that the mixed-effect predictions of quantile functions are non-decreasing, and provides shape-constrained, interpretable distributional fixed effects as in \citet{ghosal2025distributional} (Section \ref{sec:inference}). Through simulations, we demonstrate that SILDOR achieves superior estimation accuracy and predictive performance compared to the non-spatial distributional outcome regression competitor (Section \ref{sec:simulation_studies}). We apply the proposed model to the CASTNet $\tno$ concentration data, showing that spatial modeling substantially reduces posterior uncertainty and improves out-of-sample predictive performance (Section \ref{sec:real app}). We acknowledge that the proposed model has limitations and can be extended to more flexible models (Section \ref{sec:discussion}).

%The remainder of the paper is organized as follows. Section 2 presents the modeling framework, the SILDOR model, and the estimation framework. Section 3 illustrates the projected-posterior approach for inference. Section 4 reports the results from simulation studies. Section 5 presents the results from the motivating $\tno$ data application. Section 6 concludes with a discussion of contributions, limitations, and potential extensions of the proposed work.

\section{Methods} \label{sec:methods}

\subsection{Data Structure and Distributional Representations}

We consider data collected at $n$ spatial locations (sites), indexed $i = 1, \ldots, n$. Let $\mathbf{s}_i \in \mathcal{D} \subset \mathbb{R}^2$ denote the geographic coordinates of site $i$. At each site, observations are made at $J_i$ longitudinal time points (e.g., calendar months in our application), indexed $j = 1, \ldots, J_i$, with $T_{ij}$ denoting the time of the $j$th observation at site $i$. For notational simplicity, we assume a common temporal grid $\mathcal{T}=\{T_1, \ldots, T_J\}$ across sites, although the framework accommodates site-specific and unbalanced time grids without modification. At each site and longitudinal follow-up combination $(i,j)$, we observe $L_{ij}$ continuous measurements $\{Y_{ijk}\}_{k=1}^{L_{ij}}$ (e.g., the $\tno$ measurements), that we treat as a sample from a latent site and time-specific distribution with cumulative distribution function $F_{ij}$ \citep{yang2020quantile,ghosal2025distributional}.

We encode the site and time-specific distribution of such a collection of measurements through its quantile function \citep{yang2020quantile,ghosal2023distributional,ghosal2025distributional}
\begin{equation}
  Q_{ij}(p) \;=\; F_{ij}^{-1}(p) \;=\; \inf\{ y : F_{ij}(y) \ge p \},
  \qquad p \in [0,1].
  \label{eq:quantile_def}
\end{equation}

The quantile function is particularly attractive as a distributional representation since it has several desirable mathematical properties  \citep{powley2013quantile,ghosal2023distributional}, without needing any selection of a smoothing parameter as in the case of density estimation. The quantile function lies in a convex subset of $L^2[0,1]$, the squared $L_2$ distance between two quantile functions equals the squared $2$-Wasserstein distance between the corresponding distributions \citep{bigot2018upper}, and the only shape restriction it must satisfy is monotonicity (non-decreasingness). We estimate the unobserved $Q_{ij}(\cdot)$ using the empirical quantile function based on the observed measurements.
Letting $Y_{(ij,1)} \le Y_{(ij,2)} \le \cdots \le Y_{(ij,L_{ij})}$ denote the
order statistics of the $L_{ij}$ measurements, the site and longitudinal time-specific empirical quantile function \citep{hyndman1996sample} is calculated as
\begin{equation}
  \tilde{Q}_{ij}(p) \;=\; (1-w)\, Y_{(ij,\lfloor (L_{ij}+1)p \rfloor)}
        \;+\; w \, Y_{(ij,\lfloor (L_{ij}+1)p \rfloor + 1)},
  \label{eq:empquant}
\end{equation}
where $w \in [0,1)$ satisfies $(L_{ij}+1)p = \lfloor (L_{ij}+1)p \rfloor + w$, and $p \in [\frac{1}{L_{ij}+1},\frac{L_{ij}}{L_{ij}+1}]$ . When $L_{ij}$ is moderately large, $\tilde{Q}_{ij}(\cdot)$ serves as a satisfactory estimate of $Q_{ij}(\cdot)$. In our application, each station-month distribution is constructed by pooling the weekly nitrate measurements across the ten study years (2010-2019), yielding roughly $L_{ij} \approx 40$-$50$ observations for calculating the empirical
quantile function. Pooling weekly measurements across years to characterize each station-month long-term distribution is supported by the finding of \citet{ghosh2010spatio} that the chemical and meteorological covariates exhibit little year-to-year variation in their monthly medians. These site and month-specific empirical quantile functions $\tilde{Q}_{ij}(p)$ are non-decreasing over $p$, and provide more information compared to aggregated site and month-specific summaries such as the sample mean, maximum or minimum, or other sample moments based on $Y_{ijk}$s. In addition to the distributional outcome, each site and longitudinal observation $(i,j)$ carries a $q$-dimensional vector of scalar covariates $\mathbf{z}_{ij} = (z_{ij1},
\ldots, z_{ijq})^{\!\top}$. In the motivating data application, these are the monthly average
temperature and monthly total precipitation at site $i$, month $j$. Each of these is rescaled to $[0,1]$ by a monotone linear transformation so that $\mathbf{z}_{ij} \in [0,1]^q$ without loss of generality \citep{ghosal2025distributional}. The observed quantile functions are generally evaluated on a common grid $\mathcal{P} = \{p_1, p_2, \ldots, p_S\} \subset [0,1]$ of length $S$, yielding the $S$-vector $\tilde{\mathbf{Q}}_{ij} = (\tilde{Q}_{ij}(p_1), \ldots, \tilde{Q}_{ij}(p_S))^T$ for each $(i,j)$.

\subsection{Spatially-Indexed Longitudinal Distributional Outcome Regression}
We propose the spatially-indexed longitudinal distributional outcome regression (SILDOR) model, which relates the latent distributional outcome $Q_{ij}(p)$ to the scalar covariates through distributional fixed effects varying over $p$, and accounts for residual spatio-temporal dependence through a structured random-effect process $W_{ij}(p)$. Specifically, the model is given by,
\begin{equation}
  Q_{ij}(p) \;=\; \beta_0(p) \;+\; \sum_{l=1}^{q} z_{ijl}\,\beta_l(p)
              \;+\; W_{ij}(p) \;+\; \epsilon_{ij}(p),
  \qquad p \in [0,1].
  \label{eq:model}
\end{equation}
Here $\beta_0(p)$ is a distributional intercept, and $\beta_l(p)$ is the distributional fixed effect of the $l$th covariate $z_{ijl}$ at quantile level $p$. We denote by $W_{ij}(p)$ a mean-zero spatio-temporal random process that captures distributional deviations from the global mean specific to site $i$ and longitudinal time $j$. We assume that the random-effect process $W_{ij}(p)$ is independent of the covariates $\mathbf{z}_{ij}$, and $\mathbb{E}\{W_{ij}(p)\} = 0$ for all
$p \in [0,1]$. We further assume that $\epsilon_{ij}(p)$ is a mean-zero residual measurement error process independent of $W_{ij}(p)$ and satisfying $\mathbb{E}\{\epsilon_{ij}(p) \mid \mathbf{z}_{ij}, W_{ij}\} = 0$. The conditional mean is given by
\begin{equation}
  \mu_{ij}(p)
  \;\equiv\; \mathbb{E}\{ Q_{ij}(p) \mid \mathbf{z}_{ij}, W_{ij} \}
  \;=\; \beta_0(p) + \sum_{l=1}^q z_{ijl}\,\beta_l(p) + W_{ij}(p).
  \label{eq:condmean1}
\end{equation}
Under the squared Wasserstein loss (equivalently squared $L_2$ loss), $\mu_{ij}(\cdot)$ is the Wasserstein barycenter of the conditional distribution of $Q_{ij}(\cdot)$ \citep{bigot2018upper} and also the predicted quantile function. The strict-exogeneity assumption $W_{ij}(\cdot) \perp \mathbf{z}_{ij}$ ensures that the distributional fixed effects retain their interpretation as covariate effects on the marginal outcome
distribution, $\mathbb{E}\{Q_{ij}(p)\mid \mathbf{z}_{ij}\} = \beta_0(p) +
\sum_{l=1}^q z_{ijl}\beta_l(p)$. For $\mu_{ij}(\cdot)$ to be a valid quantile function, it must be non-decreasing in $p$. We first state conditions ensuring that the distributional fixed effects yield a valid quantile function across the entire covariate domain.  %This is obtained as a direct application of the joint-monotonicity result of \citet{ghosal2025distributional}.

\begin{theorem}[Monotonicity of the fixed effects prediction]
\label{thm:monotone1}
Define the fixed-effect prediction
$\mu^{\mathrm{F}}(p; \mathbf{z}) = \beta_0(p) + \sum_{l=1}^{q} z_l\,\beta_l(p)$. Then $\mu^{\mathrm{F}}(p; \mathbf{z})$ is non-decreasing in $p$ for every covariate configuration $\mathbf{z} \in [0,1]^q$ if and only if:
\begin{enumerate}[label=\textup{(\roman*)}, noitemsep]
  \item the distributional intercept $\beta_0(p)$ is non-decreasing in $p$.
  \item for every subset $\mathcal{R} \subseteq \{1,\ldots,q\}$, the additive combination $\beta_0(p) + \sum_{l \in \mathcal{R}} \beta_l(p)$ is non-decreasing in $p$.
\end{enumerate}
\end{theorem}
A proof is given in Appendix A of the supplementary material and directly follows from the joint-monotonicity result in \citet{ghosal2025distributional}. Crucially, conditions~(i)-(ii) do not require any individual  slope function $\beta_l(p)$ to be monotone; only the additive combinations need to be, providing flexibility in modeling non-monotone covariate effects \citep{ghosal2025distributional}. Theorem~\ref{thm:monotone1} governs the distributional fixed effects prediction. However, the full predicted quantile function $\mu_{ij}(p)$ in \eqref{eq:condmean1} additionally includes the site and time-specific random effect $W_{ij}(p)$. Requiring $W_{ij}(p)$ to be monotone on its own is overly restrictive and not necessary. Since, $\mu_{ij}(\cdot)$ is the predicted quantile function under SILDOR \eqref{eq:condmean1}, we enforce that the full predicted quantile function is non-decreasing, jointly across the fixed and random parts.  

\begin{proposition}[Monotonicity of the mixed-effect prediction]
\label{prop:mixed}
Let the distributional fixed effects satisfy conditions \textup{(i)-(ii)} of Theorem~\ref{thm:monotone1}. Then the predicted quantile function $\mu_{ij}(p) = \beta_0(p) + \sum_{l=1}^q z_{ijl}\beta_l(p) + W_{ij}(p)$ is non-decreasing in $p$ for every covariate configuration $\mathbf{z} \in [0,1]^q$ if and only if, for every subset $\mathcal{R} \subseteq \{1,\ldots,q\}$,
\begin{equation}
  \beta_0(p) + \sum_{l \in \mathcal{R}} \beta_l(p) + W_{ij}(p)
  \quad\text{is non-decreasing in } p .
  \label{eq:mixed_mono1}
\end{equation}
\end{proposition}

The proof is given in Appendix A of the supplementary material. In particular, $\mu_{ij}(p;\mathbf{z})$ can again be written as a convex combination of the $2^q$ vertex functions in \eqref{eq:mixed_mono1}, each of which is non-decreasing. The above condition ensures that the mixed effect prediction of the site and time-specific quantile function is monotone across the entire covariate domain, and across all spatial locations and longitudinal time-points.

\subsection{Estimation}
\subsubsection*{Basis Expansions for Fixed Effects} We model each distributional coefficient $\beta_l(p)$, $l = 0,\ldots,q$, using a Bernstein polynomial basis expansion of degree $N$: 
\begin{equation}
  \beta_l(p) \;=\; \sum_{k=0}^{N} \beta_{lk}\, b_k(p; N),
  \qquad
  b_k(p; N) = \binom{N}{k} p^k (1-p)^{N-k},
  \qquad p \in [0,1],
  \label{eq:bernstein_FE}
\end{equation}
with basis-coefficient vector $\bm{\beta}_l = (\beta_{l0}, \ldots, \beta_{lN})^{\!\top} \in \mathbb{R}^{N+1}$. 
The Bernstein basis functions are non-negative and sum to unity, i.e., $\sum_{k=0}^N b_k(p;N) = 1$ for all $p \in [0,1]$, and enjoy attractive shape-preserving properties \citep{carnicer1993shape}, particularly in the context of distributional regression  \citep{ghosal2023shape,ghosal2025distributional}. Let $\mathbf{b}_N(p)^{\!\top} = \big( b_0(p;N), b_1(p;N), \ldots, b_N(p;N) \big)$ denote the row vector of basis functions evaluated at a single
quantile level $p$. Then $\beta_l(p) = \mathbf{b}_N(p)^{\!\top}\bm{\beta}_l$. For observation $(i,j)$ corresponding to site $i$ and longitudinal time $j$ with covariate vector $\mathbf{z}_{ij} = (z_{ij1}, \ldots, z_{ijq})^{\!\top}$, we denote the covariate-scaled basis row vectors as,
\begin{equation}
  \mathbf{Z}_{ij,0}^{\!\top}(p) = \mathbf{b}_N(p)^{\!\top},
  \qquad
  \mathbf{Z}_{ij,l}^{\!\top}(p) = z_{ijl}\,\mathbf{b}_N(p)^{\!\top},
  \qquad l = 1, \ldots, q.
  \label{eq:Zrow}
\end{equation}
Now the fixed-effects contribution to $Q_{ij}(p)$ at a single quantile level $p$ can be written as:

\begin{eqnarray}
    \mu^{\mathrm{F}}(p; \mathbf{z_{ij}}) = \beta_0(p) + \sum_{l=1}^{q}  z_{ijl}\,\beta_l(p) 
=\mathbf{b}_N(p)^{\!\top}\bm{\beta}_0+\sum_{l=1}^{q} z_{ijl}\,\mathbf{b}_N(p)^{\!\top}\bm{\beta}_l \notag \\
    =\mathbf{Z}_{ij,0}^{\!\top}(p) \bm{\beta}_0+\sum_{l=1}^{q}  \mathbf{Z}_{ij,l}^{\!\top}(p)\bm{\beta}_l
    =\sum_{l=0}^q \mathbf{Z}_{ij,l}^{\!\top}(p)\,\bm{\beta}_l.
    \label{eq:Zrow1}
\end{eqnarray}
Since every empirical quantile function is evaluated on the common grid $\mathcal{P} = \{p_1,\ldots,p_S\}$,  stacking \eqref{eq:Zrow1} over $\mathcal{P}$ gives the $S \times (N+1)$ design block
\begin{equation}
  \mathbf{B}_{ij,l}
  \;=\; \big( \mathbf{Z}_{ij,l}(p_1),\, \mathbf{Z}_{ij,l}(p_2),\, \ldots,\,
  \mathbf{Z}_{ij,l}(p_S) \big)^{\!\top},
  \qquad l = 0, 1, \ldots, q.
  \label{eq:Bblock}
\end{equation}
The fixed-effects contribution to the stacked quantile vector $\tilde{\mathbf{Q}}_{ij}$ is then given by $\sum_{l=0}^q \mathbf{B}_{ij,l}\,\bm{\beta}_l$.

\subsubsection*{Spatio-Temporal Random Effects}
\label{subsec:RE}
We model the spatio-temporally correlated random effects process $W_{ij}(p)$, that capture distributional deviations from the global fixed effects nonparametrically using a tensor product basis expansion. First, $W_{ij}(p)$ is modeled as,
\begin{equation}
    W_{ij}(p) = \sum_{m=1}^{M} \psi_m(p) \, \alpha_{ijm},
    \label{eq:RE_expansion}
\end{equation}
where $\{\psi_m(p)\}_{m=1}^M$ are again taken to be Bernstein basis functions in the quantile direction. In particular, these are equivalent to Beta density basis evaluated at $p$, i.e., $\psi_m(p) = \text{Beta}(p; m, M-m+1)$ (up to normalization constant), and the coefficients $\alpha_{ijm}$ inherit the spatial and temporal dynamics for basis component $m$. Each coefficient $\alpha_{ijm}$ is further decomposed through a time-varying basis expansion:
\begin{equation}
    \alpha_{ijm} = \sum_{d=1}^{D_m} \eta_{imd} \, \phi_{md}(T_j),
    \label{eq:RE_time}
\end{equation}
where $\{\phi_{md}(t)\}_{d=1}^{D_m}$ are taken to be cubic $B$-spline basis functions evaluated at time $T_j$, and $\eta_{imd}$ are site-specific basis coefficients. Thus, we have the tensor product decomposition,
\begin{equation}
    W_{ij}(p) = \sum_{m=1}^{M}\sum_{d=1}^{D_m} \eta_{imd}\,\psi_m(p)\,\phi_{md}(T_j).
\end{equation}

Denote $\bm{\eta}_i = \text{vec}(\{\eta_{imd}\}_{m,d})$ as the $D_{\text{tot}} = \sum_{m=1}^M D_m$ dimensional vector of random effect coefficients for site $i$.  The random effect contribution to the stacked quantile vector
$\tilde{\mathbf{Q}}_{ij}$ is then given by $\bm{\Phi}_{ij}\bm{\eta}_i$, where $\bm{\Phi}_{ij}$ is an $S \times D_{\text{tot}}$ design matrix formed by the stacking the tensor product elements $\{\psi_m(p_s)\phi_{md}(T_j)\}_{m=1,d=1}^{M,d_m}$ across $s = 1,\ldots,S$.

\subsubsection*{Spatial dependence}
Spatial correlation is induced through a Gaussian-process prior on the site-level coefficient vectors $\bm{\eta}_i$.  Specifically, letting $\bm{\eta} = (\bm{\eta}_1^T, \ldots, \bm{\eta}_n^T)^T$, we assume
%Stacking $\bm{\eta} = (\bm{\eta}_1^{\!\top},
%\ldots, \bm{\eta}_n^{\!\top})^{\!\top}$, we assume the Kronecker-separable
%Gaussian prior
\begin{equation}
  \bm{\eta} \;\sim\;
  \mathcal{N}\!\big( \mathbf{0},\; \bm{\Sigma}_s \otimes \bm{\Lambda} \big),
  \qquad
  \bm{\Lambda} = \mathrm{diag}(\lambda_1, \ldots, \lambda_{D_{\mathrm{tot}}}),
  \label{eq:GP_prior1}
\end{equation}
where $\bm{\Sigma}_s$ is an $n \times n$ spatial correlation matrix and
$\mathbf{\Lambda} = \text{diag}(\lambda_1, \ldots, \lambda_{D_{\text{tot}}})$ is a diagonal variance matrix that allows different scaling across the basis components. $\bm{\Sigma}_s$ is parameterized as
\begin{equation}
  [\bm{\Sigma}_s]_{ii'}
  = r\,C\!\big( \| \mathbf{s}_i - \mathbf{s}_{i'} \|;\,
        \rho_s, \nu \big)
  + (1-r)\,\mathbb{1}(i = i'),
  \label{eq:matern}
\end{equation}
where $C(\cdot;\rho_s,\nu)$ is the Mat\'ern correlation function with range parameter $\rho_s > 0$ and smoothness parameter $\nu > 0$, and $r \in (0,1)$ is the ratio of partial sill to sill.

\subsubsection*{Conditional model}
Conditionally on the random effects, model (\ref{eq:model}) for the stacked quantile vector is reformulated based on the basis expansions as,
\begin{equation}
  \tilde{\mathbf{Q}}_{ij}
  \;=\; \sum_{l=0}^{q} \mathbf{B}_{ij,l}\,\bm{\beta}_l
        \;+\; \bm{\Phi}_{ij}\,\bm{\eta}_i
        \;+\; \bm{\epsilon}_{ij},
  \label{eq:model_matrix1}
\end{equation}
where the error $\bm{\epsilon}_{ij} \sim \mathcal{N}(\mathbf{0}, \sigma_\epsilon^2 \mathbf{I}_S)$ captures residual variation not captured by the fixed or structured random effect. Stacking \eqref{eq:model_matrix1} over all site-longitudinal combination $(i,j)$ we have
\begin{equation}
 \mathbf{Q} = \mathbf{X}\bm{\beta} + \bm{\Phi}\bm{\eta} + \bm{\epsilon},   
\end{equation}
where $\mathbf{Q} = (\tilde{\mathbf{Q}}_{11}^{\!\top}, \ldots, \tilde{\mathbf{Q}}_{1J}^{\!\top}, \ldots, \tilde{\mathbf{Q}}_{n1}^{\!\top}, \ldots, \tilde{\mathbf{Q}}_{nJ}^{\!\top})^{\!\top} \in \mathbb{R}^{nJS}$. We denote by $\mathbf{X}$ the matrix obtained by stacking $\mathbf{X}_{ij}$ over $(i,j)$, where the fixed-effect design blocks $\mathbf{X}_{ij} = (\mathbf{B}_{ij,0}, \ldots, \mathbf{B}_{ij,q})$. The fixed-effect coefficient vector is denoted by $\bm{\beta} = (\bm{\beta}_0^{\!\top}, \ldots, \bm{\beta}_q^{\!\top})^{\!\top}$. $\bm{\Phi} = \mathrm{blockdiag}(\bm{\Phi}_1, \ldots, \bm{\Phi}_n)$ is the block-diagonal random-effect design with block $\bm{\Phi}_i$ obtained by stacking $\bm{\Phi}_{ij}$ over $j$. Analogously  $\bm{\epsilon}$ denotes the stacked residual vector, with $\bm{\epsilon} \sim \mathcal{N}(\mathbf{0}, \sigma_\epsilon^2 \mathbf{I}_{nJS})$.

\section{Bayesian Inference} \label{sec:inference}
%\textcolor{red}{Double Check and edit: SM}
A key computational challenge for a Bayesian approach here is that the shape constraints of proposition 1 couple the fixed-effect coefficients $\bm{\beta}$ and the random-effect coefficients $\bm{\eta}$ through $2^q$ inequality constraints per observation, rendering a fully constrained sampler computationally expensive. We employ a two-stage projected posterior strategy \citep{lin2014bayesian, chakraborty2021convergence, chakraborty2021coverage, chakraborty2022rates,pal2025projection,pal2026bayesian} for posterior inference. We run an unconstrained Metropolis-within-Gibbs sampler that exploits the conjugate Gaussian structure of \eqref{eq:model_matrix1} and then project each posterior draw onto the monotonicity-constrained space through a sequence of small quadratic programming solvers. This delivers non-decreasing posterior summaries of both the distributional fixed effects and the subject-specific predicted quantile functions at a fraction of the cost of constrained sampling, while retaining the strong theoretical guarantees of projected posterior inference \citep{chakraborty2021convergence,chakraborty2021coverage,chakraborty2022rates}.

\subsection{Priors and the Unconstrained MCMC Sampler}
\label{subsec:mcmc}

We complete the Bayesian model specification with the following standard choices for priors, selected to be vaguely informative and conditionally conjugate where possible:

\begin{align*}
      \bm{\beta}_l &\sim  \mathcal{N}\left(\bm{0},\tau^2_\beta\bm{I}_{N+1}\right) \,,\ l = 0, \ldots q\\
      \sigma^2_\epsilon &\sim IG\left(a_\epsilon,b_\epsilon\right),\\
      \bm{\eta} \mid \bm{\Sigma}_s, \bm{\Lambda} &\sim \mathcal{N}\left(\bm{0},\, \bm{\Sigma}_s \otimes \bm{\Lambda}\right),\\
     \lambda_d &\sim IG\left(a_\lambda,b_\lambda\right) \,,\ d = 1,\ldots , D_{\mathrm{tot}},\\
      \log \rho_s &\sim \mathcal{N}\left(\mu_\rho, \sigma_\rho^2\right),\\
      \mbox{logit} (r) &\sim \mathcal{N}\left( \mu_r, \sigma_r^2 \right).
\end{align*}

 These choices ensure that the full conditional densities for $\bm{\beta}_l \,,\ l =0,\ldots, q$, $\bm{\eta}$, $\sigma^2_\epsilon$, and $\lambda_d \,,\ d=1, \ldots, D_{\mbox{tot}}$ are conjugates and easy to draw from. The full conditional densities for $\log \rho_s$ and $\mbox{logit}(r)$ do not take a known closed-form expression and therefore requires Metropolis draws from the full conditionals to sample from appropriately. 
%For each of the $B$ many samples of $\bm{\theta}^{(b)} = \left( \bm{\beta}_0^{(b) T}, \ldots, \bm{\beta}_q^{(b) T}, \bm{\eta}^{(b) T},\sigma_\epsilon^{2(b)}, \lambda_1^{(b)}, \ldots, \lambda_{D_{\mathrm{tot}}}^{(b)}, \log \rho_s^{(b)}, \mbox{logit} (r)^{(b)} \right)^T$, $b = 1, \ldots, B$, we generate $\bm{\eta}_{i'}^{(b)}$, the predicted value of the spatial random effect at location $i'$, from its conditional predictive distribution which is also Gaussian.
The full conditional distributions for the sampler are as below:

\subsection*{(a) Fixed effects $\bm{\beta}$}

Conditional on $(\bm{\eta}, \sigma_\epsilon^2)$, the full conditional is Gaussian,
\begin{equation*}
  \bm{\beta} \mid \cdot \;\sim\;
  \mathcal{N}\!\big( \mathbf{m}_\beta, \mathbf{V}_\beta \big),
  \quad
  \mathbf{V}_\beta
    = \left(\sigma_\epsilon^{-2}\,\mathbf{X}^{\!\top}\mathbf{X}
      + \tau_\beta^{-2}\,\mathbf{I}\right)^{-1},
  \quad
  \mathbf{m}_\beta
    = \sigma_\epsilon^{-2}\,\mathbf{V}_\beta\,
      \mathbf{X}^{\!\top}(\mathbf{Q} - \bm{\Phi}\bm{\eta}).
  \label{eq:fc_psi}
\end{equation*}

%The cross-product $\mathbf{X}^{\!\top}\mathbf{X}$ is precomputed once.

\subsection*{(b) Random effects $\bm{\eta}$}
Conditional on $(\bm{\beta}, \sigma_\epsilon^2, \rho_s, r, \bm{\lambda})$, and writing $\bm{\Sigma}_\eta = \bm{\Sigma}_s \otimes \bm{\Lambda}$,
\begin{equation*}
  \bm{\eta} \mid \cdot \;\sim\;
  \mathcal{N}\!\big( \mathbf{m}_\eta, \mathbf{V}_\eta \big),
  \quad
  \mathbf{V}_\eta
    = \left(\sigma_\epsilon^{-2}\,\bm{\Phi}^{\!\top}\bm{\Phi}
      + \bm{\Sigma}_{\eta}^{-1}\right)^{-1},
  \quad
  \mathbf{m}_\eta
    = \sigma_\epsilon^{-2}\,\mathbf{V}_\eta\,
      \bm{\Phi}^{\!\top}(\mathbf{Q} - \mathbf{X}\bm{\beta}).
  \label{eq:fc_eta}
\end{equation*}
The Kronecker-separable structure of the prior precision $\bm{\Sigma}_\eta^{-1} =
\bm{\Sigma}_s^{-1} \otimes \bm{\Lambda}^{-1}$ allows this term to be computed at
cost $O(n^3 + D_{\text{tot}}^3)$, rather than $O((nD_{\text{tot}})^3)$ for a
direct inversion. Draws from $\bm{\eta} \mid \cdot$ are then obtained via the
Cholesky factor of $\mathbf{V}_\eta^{-1}$, using forward and backward solves
rather than an explicit matrix inversion, further reducing computation time.

%The separability assumption on the prior variance $\boldsymbol{\Sigma}_\eta$ implies that $\boldsymbol{\Sigma}_\eta$ can be inverted in $\mathcal{O}\left( n^3 + D_{tot}^3\right)$ flops only. 
%The block diagonal structure of $\bm{\Phi}$ and the diagonal structure of $\bm{\Lambda}$ ensures a sparse structure for $\bm{V}_\eta^{-1}$ which can be leveraged for a sparse Cholesky decomposition. Appropriate usage of forward and backward solve algorithms allow for draws from this Gaussian distribution without needing to completely invert $\bm{V}_\eta$, reducing computation time.

\subsection*{(c) Error variance $\sigma_\epsilon^2$}
With $N_{\mathrm{obs}} = nJS$ the total number of scalar quantile evaluations,
\begin{equation*}
  \sigma_\epsilon^2 \mid \cdot \;\sim\;
  IG\!\Big(
    a_\epsilon + \tfrac{N_{\mathrm{obs}}}{2},\;
    b_\epsilon + \tfrac{1}{2}\,
      \| \mathbf{Q} - \mathbf{X}\bm{\beta} - \bm{\Phi}\bm{\eta} \|^2
  \Big).
  \label{eq:fc_sigma}
\end{equation*}

\subsection*{(d) Component variances $\lambda_d$}
Using the separability of the prior, each $\lambda_d$ has an independent Inverse-Gamma full conditional,
\begin{equation*}
  \lambda_d \mid \cdot \;\sim\;
  IG\!\Big(
    a_\lambda + \tfrac{n}{2},\;
    b_\lambda + \tfrac{1}{2}\,
      \bm{\eta}_{\cdot d}^{\!\top}\bm{\Sigma}_s^{-1}\bm{\eta}_{\cdot d}
  \Big),
  \qquad d = 1, \ldots, D_{\mathrm{tot}},
  \label{eq:fc_lambda}
\end{equation*}
where $\bm{\eta}_{\cdot d}$ is the $n \times 1$ vector corresponding to the $d$-th set of spatial effects for the $n$ locations. In other words, $\bm{\eta}_{\cdot d} = \left( \eta_{n(d-1) + 1}, \ldots , \eta_{nd} \right)^{\sf T}$ with $\eta_i$ being the $i$-th entry of $\bm{\eta}$.

\subsection*{(e) Spatial range $\rho_s$}

We identify $\log \rho_s$ as a parameter for the model and update the same. Samples from the posterior distribution of $\rho_s$ is in turn generated by first generating samples for $\log \rho_s$ from its posterior distribution followed by an exponentiation of them. The full conditional density for $\log \rho_s$ is of a non-standard form as below: \[\pi\left(\log \rho_s \vert \cdot \right) \simeq -\frac{D_{tot}}{2}\log \det \bm{\Sigma}_s - \frac{1}{2}\bm{\eta}^{\sf T}\bm{\Sigma}_{\eta}^{-1}\bm{\eta} - \frac{1}{2\sigma_\rho^2} \left(\log \rho_s - \mu_\rho\right)^2,\] where $\simeq$ denotes equality upto an additive constant. We employ a random walk Metropolis-Hastings (RWMH) update for $\log \rho_s$ from its full conditional distribution.

\subsection*{(f) Nugget $r$}

Samples from the posterior distribution of $r$ is similarly generated by first generating samples for $\text{logit} (r)$ from its posterior distribution followed by their inverse-logit transformation transformation $\left(h(x) = \frac{1}{1+\exp(-x)}\right)$. The full conditional density for $\text{logit} (r)$ is of a non-standard form as below: 
\[\pi\left(\text{logit} (r) \vert \cdot \right) \simeq -\frac{D_{tot}}{2}\log \det \bm{\Sigma}_s - \frac{1}{2}\bm{\eta}^{\sf T}\bm{\Sigma}_{\eta}^{-1}\bm{\eta} - \frac{1}{2\sigma_r^2} \left(\text{logit} (r) - \mu_r\right)^2.\]
We employ a RWMH update for $\text{logit} (r)$ from its full conditional distribution.

%The sampler is run for $N_{\mathrm{iter}}$ iterations with burn-in $N_{\mathrm{burn}}$ and thinning $N_{\mathrm{thin}}$, retaining $B = (N_{\mathrm{iter}} - N_{\mathrm{burn}})/N_{\mathrm{thin}}$ draws $\{ (\bm{\beta}^{(b)}, \bm{\eta}^{(b)}, \sigma_\epsilon^{2(b)},\bm{\lambda}^{(b)}, \rho_s^{(b)}, r^{(b)}) \}_{b=1}^B$. %Convergence is monitored by trace plots and effective sample sizes of the distributional coefficients and
%spatial parameters.

\subsection{Projected-Posterior Inference for the Fixed Effects}
\label{subsec:proj_fixed}
The unconstrained sampler of Section~\ref{subsec:mcmc} ignores the monotonicity
constraints of Theorem~\ref{thm:monotone1} and proposition 1.  Based on the Bernstein-basis expansions of $\beta_j(p)$, the conditions in Theorem~\ref{thm:monotone1} reduce to linear inequality constraints on $\bm{\beta} = (\bm{\beta}_0^T, \bm{\beta}_1^T, \ldots, \bm{\beta}_q^T)^T$ of the form $\mathbf{A}\bm{\beta} \geq \mathbf{0}$, where $\mathbf{A}$ is a block-structured constraint matrix \citep{ghosal2023shape, ghosal2025distributional}. $\mathbf{A}$ is formed by stacking the first-difference matrices $\mathbf{A}_1$, as illustrated in Appendix~A of the supplementary material. To obtain valid inference for the distributional fixed effects which adhere to the shape constraints, we project each posterior draw $\bm{\beta}^{(b)}$
onto the constraint set
$\mathcal{C}_\beta = \{ \bm{\beta} : \mathbf{A}\bm{\beta} \ge \mathbf{0} \}$, under the least squares loss function corresponding to the fixed effect prediction.
\begin{equation}
  \widehat{\bm{\beta}}^{(b)}
  \;=\; \underset{\bm{\beta} \in \mathcal{C}_\beta}{\arg\min}\;
        \big\| \mathbf{X}\bm{\beta} - \mathbf{X}\bm{\beta}^{(b)} \big\|_2^2 .
  \label{eq:proj_psi}
\end{equation}
This quadratic program can be efficiently solved by the dual active-set algorithm of
\citet{goldfarb1983numerically} and is implemented using \texttt{quadprog} in R.  The projected draws $\{\widehat{\bm{\beta}}^{(b)}\}_{b=1}^B$ constitute a sample from the
projected posterior of the constrained fixed effects. Point estimates of the
distributional coefficients are then obtained as the projected posterior means,
$\widehat{\beta}_r(p) = B^{-1} \sum_{b=1}^B \widehat{\beta}_r^{(b)}(p)$. 
The $100(1-\alpha)\%$ pointwise credible intervals are also obtained from the empirical quantiles of $\{ \widehat{\beta}_r^{(b)}(p) \}_b$ across the projected posterior draws.

\subsection{Projected-Posterior Mixed-effect  Prediction of Distributional Outcomes}
\label{subsec:proj_RE}
Valid prediction of the site and time-specific quantile functions
$\mu_{ij}(\cdot)$ requires the joint monotonicity of
Proposition~\ref{prop:mixed}. Given the projected fixed-effect estimate, we constrain the
random-effect loadings so that the full predicted quantile function is non-decreasing for every additive covariate combination. Note that this does not require $W_{ij}(p)$ to
be monotone in isolation. In particular, given the projected fixed-effect draw $\widehat{\bm{\beta}}^{(b)}$ from the previous step and the
random-effect draw $\bm{\eta}^{(b)}$ from the unconstrained MCMC sampler, we  reconstruct, for each $(i,j)$, the
unconstrained loadings
$\bm{\alpha}_{ij}^{(b)} = (\alpha_{ij1}^{(b)}, \ldots, \alpha_{ijM}^{(b)})^{\!\top}$
via $\alpha_{ijm}^{(b)} = \sum_{d=1}^{D_m} \eta_{imd}^{(b)}\,\phi_{md}(T_j)$. Then we obtain unconstrained draws of the spatio-temporal random process as $W_{ij}^{(b)}(p) = \sum_{m=1}^M \psi_m(p)\,\alpha_{ijm}^{(b)}$. Finally, we solve
the projection problem, for each $(i,j)$,
\begin{equation}
  \widehat{\bm{\alpha}}_{ij}^{(b)}
  \;=\; \underset{\bm{\alpha} \in \mathbb{R}^M}{\arg\min}\;
        \big\| \bm{\Psi}\bm{\alpha} - \bm{\Psi}\bm{\alpha}_{ij}^{(b)} \big\|_2^2
  \quad \text{subject to} \quad
  \mathbf{A}_{\mathrm{RE}}\, \bm{\alpha} \;\ge\; \mathbf{b}_{ij}^{(b)}.
  \label{eq:proj_RE}
\end{equation}
Here $\bm{\Psi}$ is the $S \times M$ quantile-basis matrix with entries
$\psi_m(p_s)$. The constraint encodes Proposition~\ref{prop:mixed}. Let $\widetilde{\mathbf{A}}_1 \in \mathbb{R}^{(M-1)\times M}$ denote the Bernstein first-difference matrix (see supplementary material) corresponding to the random-effect Bernstein basis. We denote by
$\mathbf{A}_1^{\mathrm{FE}} \in \mathbb{R}^{N \times (N+1)}$ the
first-difference matrix on the degree-$N$ fixed-effect Bernstein basis. The joint monotonicity constraint for combination $\mathcal{R}$ is $\widetilde{\mathbf{A}}_1\bm{\alpha} \geq \widetilde{\mathbf{b}}_{ij,\mathcal{R}}^{(b)}$,
where $\widetilde{\mathbf{b}}_{ij,\mathcal{R}}^{(b)} \in \mathbb{R}^{M-1}$
is the first $(M-1)$ entries of $-\mathbf{A}_1^{\mathrm{FE}}\big(\widehat{\bm{\beta}}_0^{(b)} + \sum_{r\in\mathcal{R}} z_{ijr}\widehat{\bm{\beta}}_r^{(b)}\big)$,
corresponding to the active (non-trivial) difference constraints of the random effect basis.  Stacking over all $2^q$ subsets $\mathcal{R}$, we get
\begin{equation}
  \mathbf{A}_{\mathrm{RE}} = \mathbf{1}_{2^q} \otimes \widetilde{\mathbf{A}}_1
  \;\in\; \mathbb{R}^{2^q(M-1)\times M},
  \qquad
  \mathbf{b}_{ij}^{(b)} =
  \Big(\widetilde{\mathbf{b}}_{ij,\mathcal{R}}^{(b)}\Big)_{\mathcal{R}\subseteq\{1,\ldots,q\}}
  \;\in\; \mathbb{R}^{2^q(M-1)},
\end{equation}
leading to a QP with $M$ variables and $2^q(M-1)$ active constraints. For example, for the  $q=2$ case (as in our application), we have four blocks, $ \mathbf{A}_{\mathrm{RE}}
  = \mathbf{1}_{4} \otimes \widetilde{\mathbf{A}}_1$, and we have  
\newpage
\begin{equation}
  \mathbf{b}_{ij}^{(b)}
  = -\begin{pmatrix}
      \big[\mathbf{A}_1^{\mathrm{FE}}\,\widehat{\bm{\beta}}^{(b)}_{0}\big]_{1:(M-1)}\\[6pt]
      \big[\mathbf{A}_1^{\mathrm{FE}}\,(\widehat{\bm{\beta}}^{(b)}_{0}
           +z_{ij1}\widehat{\bm{\beta}}^{(b)}_{1})\big]_{1:(M-1)}\\[6pt]
      \big[\mathbf{A}_1^{\mathrm{FE}}\,(\widehat{\bm{\beta}}^{(b)}_{0}
           +z_{ij2}\widehat{\bm{\beta}}^{(b)}_{2})\big]_{1:(M-1)}\\[6pt]
      \big[\mathbf{A}_1^{\mathrm{FE}}\,(\widehat{\bm{\beta}}^{(b)}_{0}
           +z_{ij1}\widehat{\bm{\beta}}^{(b)}_{1}
           +z_{ij2}\widehat{\bm{\beta}}^{(b)}_{2})\big]_{1:(M-1)}
    \end{pmatrix}
  \in \mathbb{R}^{4(M-1)}.
  \label{eq:ARE}
\end{equation}
where $[\cdot]_{1:(M-1)}$ denotes the first $(M-1)$ entries of the
enclosed vector, retaining only the active difference constraints
corresponding to the degree $(M-1)$ random effect Bernstein basis. Each block in $\mathbf{b}_{ij}^{(b)}$ corresponds to the intercept-only, intercept-plus-temperature,
intercept-plus-precipitation, and full additive combinations, respectively. Note that the left-hand constraint matrix $ \widetilde{\mathbf{A}}_1$ is identical across
blocks. Only the right-hand
side $\mathbf{b}_{ij}^{(b)}$ varies by combination and by observation $(i,j)$
through $\mathbf{z}_{ij}$. Imposing all $2^q(M-1)$ constraints simultaneously
restricts $\bm{\alpha}$ to the intersection of the corresponding feasible
half-spaces, so the single projected loading vector
$\widehat{\bm{\alpha}}_{ij}^{(b)}$ produces a valid quantile function across the
entire covariate hypercube $[0,1]^q$. Each projection \eqref{eq:proj_RE} is a QP of
dimension $M$ (small), and the $nJ$ per-observation QPs are embarrassingly
parallel, and thus the whole algorithm is computationally efficient. The
resulting constrained prediction
\begin{equation}
  \widehat{Q}_{ij}^{(b)}(p)
  \;=\; \widehat{\beta}_0^{(b)}(p)+\sum_{r=1}^{q} z_{ijr}\,\widehat{\beta}_r^{(b)}(p)
        \;+\; \sum_{m=1}^{M} \psi_m(p)\,\widehat{\alpha}_{ijm}^{(b)}
  \label{eq:pred_constrained}
\end{equation}
is then a valid (non-decreasing) quantile function by construction. The
posterior predictive mean quantile function is
$\widehat{Q}_{ij}(p) = B^{-1} \sum_{b=1}^B \widehat{Q}_{ij}^{(b)}(p)$, and
predictive credible bands follow from the empirical quantiles of
$\{ \widehat{Q}_{ij}^{(b)}(p) \}_{b=1}^{B}$.

\subsection{Prediction at a New Spatial Location}
For each of the $B$ many samples of $\bm{\theta}^{(b)} = \left( \bm{\beta}_0^{(b) T}, \ldots, \bm{\beta}_q^{(b) T}, \bm{\eta}^{(b) T},\sigma_\epsilon^{2(b)}, \lambda_1^{(b)}, \ldots, \lambda_{D_{\mathrm{tot}}}^{(b)}, \log \rho_s^{(b)}, \mbox{logit} (r)^{(b)} \right)^T$, $b = 1, \ldots, B$, we generate $\bm{\eta}_{i'}^{(b)}$, the predicted value of the spatial random effect at location $i'$, from its conditional prior predictive distribution which is also Gaussian. \[\bm{\eta}_{pred} \sim \mathcal{N}\left( \bm{\Sigma}_{p, \eta}\bm{\Sigma}_\eta^{-1}\bm{\eta}, \bm{\Sigma}_{p} - \bm{\Sigma}_{p,\eta}\bm{\Sigma}_{\eta}^{-1}\bm{\Sigma}_{\eta,p} \right),\] where $\bm{\Sigma}_{p} = \bm{\Sigma}_{pred} \otimes \bm{\Lambda}$, $\bm{\Sigma}_{\eta, p} = \bm{\Sigma}_{s,pred} \otimes \bm{\Lambda}$ and $\bm{\Sigma}_{p,\eta} = \bm{\Sigma}_{\eta,p}^T$ with $\left(\bm{\Sigma}_{pred}\right)_{ij} = rC(\mathbf{s}^*_{i},\mathbf{s}^*_j; \rho_s, \nu) + (1-r)\mathbb{1}(i=j)$ and $\left(\bm{\Sigma}_{s,pred}\right)_{ij} = rC(\mathbf{s}_{i},\mathbf{s}^*_j; \rho_s, \nu) + (1-r)\mathbb{1}(\mathbf{s}^*_i = \mathbf{s}_j)$ where $\mathbf{s}^*$ is a prediction location and $\mathbf{s}$ is an observation location and $C(\cdot; \rho, \nu)$ is a Mat\'{e}rn correlation kernel with spatial range parameter $\rho$ and smoothness parameter $\nu$.

\subsection{Choice of Hyperparameters and Tuning Parameters}
\label{choice}
In all analyses, we set $\tau_\beta^2 = 10^4$, $a_\epsilon = b_\epsilon = a_\lambda = b_\lambda = 0.1$, $\mu_\rho = \mu_r = 0$, and $\sigma_\rho^2 = \sigma_r^2 = 10^2$  yielding diffuse priors on all the components. The number of Bernstein basis polynomials for the fixed effects $N$,
and the number of basis functions used for the spatio-temporal random effects ($M,D_m$) control the smoothness of these functions. We follow a truncated basis approach \citep{fan2015functional} and choose a moderate number of basis functions using a Deviance Information Criterion (DIC) in our data analysis. In particular, we set  $\mathrm{DIC} = \overline{D(\bm{\theta})} + p_D$, where
$D(\bm{\theta}) = -2 \log p(\mathbf{Q} \mid \bm{\theta})$ is the deviance under
the Gaussian working likelihood under \eqref{eq:model_matrix1}, $\overline{D}$ is its
posterior mean, and the effective number of parameters is
$p_D = \overline{D(\bm{\theta})} - D(\overline{\bm{\theta}})$, where
$\overline{\bm{\theta}}$ denotes the posterior mean of $(\bm{\beta}, \bm{\eta},
\sigma_\epsilon^2)$.

%\subsection{Markov Chain Monte Carlo}
%\subsection{Posterior Summaries}
%\section{Simulation Studies}  

\section{Simulation Studies}
\label{sec:simulation_studies}

We use numerical simulations to evaluate the performance of the proposed SILDOR framework. We consider the following simulation design.

\subsection{Data Generating Mechanism}
\label{subsec:sim_dgp}

We generate spatially-indexed longitudinal distributional outcomes from the
SILDOR model~\eqref{eq:model} with $q=2$ scalar covariates, mimicking the
structure of our motivating $\tno$ application. In particular, the model is given by,
\begin{equation}
  Q_{ij}(p) = \beta_0(p) + \sum_{l=1}^{2} z_{ijl}\beta_l(p)
              + W_{ij}(p) + \epsilon_{ij}(p),  \quad p \in [0,1], 1\le i\le n, 1\le j \le J.
  \label{eq:modelsim}
\end{equation}
We consider $n$ spatial
locations arranged on a regular $\sqrt{n} \times \sqrt{n}$ lattice over $[0,1]^2$. We vary $n \in \{25, 49, 100\}$, and $J = 12$ time
points are considered for each spatial location.  Each
site-time combination $(i,j)$ has two covariates $z_{ij1}, z_{ij2}$, each generated independently from a $\mathrm{U}(0,1)$ distribution. The true distributional fixed effects are specified as
$\beta_0(p) = 5 + 10p + 5p^2, 
  \beta_1(p) = p^3\,(6p^2 - 15p + 10),
  \beta_2(p) = \sin\!\big(\tfrac{\pi}{2}p\big).
  \label{eq:sim_betas}$
The spatio-temporal random effect $W_{ij}(p)$ is generated from the
tensor-product decomposition similar to equation (12). We use $M = 3$ Bernstein basis functions for $\psi_m(p)$, and in
the time direction we use
$D = 4$ B-spline basis functions per component, where component $m$ have spline degree $m+1$ for $\phi_{md}(\cdot)$. The site specific loadings $\bm{\eta}_i \in
\mathbb{R}^{D_{\mathrm{tot}}}$, with $D_{\mathrm{tot}} = MD = 12$ are stacked
to $\boldsymbol{\eta}$ with $\bm{\eta} \sim \mathcal{N}(\mathbf{0}, \bm{\Sigma}_s \otimes \bm{\Lambda})$, where $\bm{\Sigma}_s$ is a Mat\'ern correlation matrix given in ~\eqref{eq:matern},
with range $\rho_s = 0.1$, smoothness $\nu = 3/2$, and signal-to-noise ratio
$r = 0.9$. The matrix $\bm{\Lambda} =
\mathrm{diag}(\lambda_1, \ldots, \lambda_{D_{\mathrm{tot}}})$ has decreasing
diagonal entries $\lambda_d$ equally spaced between $0.2$ and $0.05$. The residual vector $\bm{\epsilon}_{ij} \sim \mathcal{N}(\mathbf{0}, \sigma_\epsilon^2 \mathbf{I}_S)$, with $\sigma_\epsilon = 0.1$. Given the generated fixed and random effects, the observed quantile vectors $\tilde{\mathbf{Q}}_{ij}$ are formed based on (\ref{eq:modelsim}), and evaluated on an equispaced grid $\mathcal{P} = \{p_1, \ldots, p_S\}$ of length $S = 21$ in $[0,1]$. To ensure that each simulated outcome is a valid (non-decreasing) quantile function, the error vector $\bm{\epsilon}_{ij}$ is
redrawn until the observed stacked quantile vector $\tilde{\mathbf{Q}}_{ij}$ is monotone on $\mathcal{P}$ using a rejection-based sampling. We generate $R=100$ replicated datasets for each simulation scenario.

\subsection{Simulation Results}
\label{subsec:sim_res}
The distributional fixed effects $\beta_r(\cdot)$,
$r = 0, 1, 2$, are modeled using Bernstein polynomials of common degree
$N = 5$. For the tensor product decomposition of the random error process $W_{ij}(p)$ as in (12) we used  $M = 3$ Bernstein basis functions $\psi_m(p)$, and $D_m=D=7$ cubic B-spline basis functions in the time direction for  $\phi_{md}(\cdot)$. Note that this implies the fitted process is misspecified, and we are evaluating the performance of SILDOR in this more general scenario. We fit the SILDOR model using the two-stage projected-posterior algorithm of
Section~\ref{sec:inference}. We benchmark the estimation performance of SILDOR against the
non-spatial distributional outcome regression (DOR) of
\citet{ghosal2025distributional}, which ignores spatial dependence. For SILDOR, we use $20,000$ iterations, $10,000$ burn-in, and thinning interval of $5$, retaining $B=2,000$ MCMC samples for each dataset. Supplementary Figure S1 displays the trace plot of parameters for a Monte-Carlo replication ($n=100$), showing successful convergence for the unconstrained sampler.

We assess the estimation performance of the distributional fixed effects, using the root integrated squared bias (R-Bias$^2$), integrated standard deviation (SD), and root mean integrated squared error (RMSE) of each $\hat{\beta}_r(\cdot)$. First, we use the mean of the projected-posterior  MCMC samples to obtain the fixed effect estimates $\hat{\beta}_r^{(j)}(\cdot)$, for each replicated dataset $j$. Then the evaluation metrics are defined for each $\beta_r(\cdot)$ as,
$
  \mathrm{RMSE}
  = \sqrt{\frac{1}{R}\sum_{j=1}^{R}\int_0^1
      \{\hat{\beta}_r^{(j)}(p) - \beta_r(p)\}^2\,dp}$. 
Here the Monte-Carlo mean estimate is given by $\bar{\hat{\beta}}_r(\cdot) = R^{-1}\sum_j
\hat{\beta}_r^{(j)}(\cdot)$. Then we define R-Bias$^2
  = \sqrt{\int_0^1
      \{\bar{\hat{\beta}}_r(p) - \beta_r(p)\}^2\,dp},
  $ and 
  $\mathrm{SD}
  = \sqrt{\frac{1}{R}\sum_{j=1}^{R}\int_0^1
      \{\hat{\beta}_r^{(j)}(p) - \bar{\hat{\beta}}_r(p)\}^2dp}$.
Table \ref{tab:sim_A1} displays these metrics across $n=25,49,100$ spatial locations for the SILDOR and the DOR method. It can be seen that, on average, the proposed SILDOR method produces $13.1$ times smaller RMSE for  $\beta_0(\cdot)$, $7$ times smaller RMSE for  $\beta_1(\cdot)$, and $7.3$ times smaller RMSE for  $\beta_2(\cdot)$, compared to the DOR method. 
Inspecting the bias-variance decomposition, the integrated squared bias is negligible relative to the integrated variance for both methods. The large RMSE gains of SILDOR over DOR are therefore driven entirely by reduction in standard deviation, reflecting the efficiency gains from borrowing strength across spatially correlated locations. The performance of SILDOR improves as $n$ increases in terms of lower RMSE, indicating consistency of the estimator.

\begin{table}[H]
\centering
\caption{Root integrated squared bias (R-Bias$^2$), standard deviation (SD), and root mean square error (RMSE) (all multiplied by 100 and rounded to 3 decimal points) of the
estimated distributional coefficients $\beta_0(p)$, $\beta_1(p)$, and
$\beta_2(p)$ averaged over 100 Monte-Carlo replications for the non-spatial (DOR) and spatial (SILDOR) models for different numbers of spatial locations ($n$).}
\label{tab:sim_A1}
\begin{tabular}{llcccccc}
\toprule
& & \multicolumn{3}{c}{SILDOR} & \multicolumn{3}{c}{DOR} \\
\cmidrule(lr){3-5} \cmidrule(lr){6-8}
Coefficient & $n$ & R-Bias$^2$ & SD & RMSE & R-Bias$^2$ & SD & RMSE \\
\midrule
\multirow{3}{*}{$\beta_0(p)$}
 & 25  & $0.080$ & $1.285$ & $1.285$
        & $1.192$ & $11.619$ & $11.662$ \\
 & 49  & $0.142$ & $0.948$ & $0.958$
        & $0.811$ & $10.954$ & $11.000$ \\
 & 100 & $0.051$ & $0.610$ & $0.612$
        & $1.616$ & $11.358$ & $11.446$ \\
\midrule
\multirow{3}{*}{$\beta_1(p)$}
 & 25  & $0.210$ & $1.520$ & $1.533$
        & $0.680$ & $10.149$ & $10.198$ \\
 & 49  & $0.150$ & $1.044$ & $1.058$
        & $0.469$ & $7.642$ & $7.655$ \\
 & 100 & $0.046$ & $0.703$ & $0.705$
        & $0.335$ & $5.070$ & $5.079$ \\
\midrule
\multirow{3}{*}{$\beta_2(p)$}
 & 25  & $0.126$ & $1.480$ & $1.487$
        & $1.473$ & $10.770$ & $10.863$ \\
 & 49  & $0.098$ & $0.978$ & $0.983$
        & $0.923$ & $7.450$ & $7.503$ \\
 & 100 & $0.065$ & $0.724$ & $0.727$
        & $0.365$ & $5.020$ & $5.040$ \\
\bottomrule
\end{tabular}
\end{table}

We also evaluate the empirical coverage of the $95\%$ projection-based credible bands for each distributional effect. Table \ref{tab:sim_coverage} reports the average pointwise coverage (over $\mathcal{P}$) of the distributional fixed effects from the SILDOR method. We observe that across all scenarios, SILDOR provides near-nominal coverage for all three distributional effects.

\begin{table}[H]
\centering
\caption{Average pointwise coverage of the $95\%$ projection-based credible
bands for the estimated distributional coefficients $\beta_0(p)$, $\beta_1(p)$,
and $\beta_2(p)$ over 100 Monte-Carlo replications using the SILDOR method for different number of observation locations ($n$).}
\label{tab:sim_coverage}
\begin{tabular}{lccc}
\toprule
$n$ & $\beta_0(p)$ & $\beta_1(p)$ & $\beta_2(p)$ \\
\midrule
25  &0.95 &0.93 &0.93 \\
49  &0.93 &0.93 &0.94 \\
100 &0.95 &0.93 &0.93 \\
\bottomrule
\end{tabular}
\end{table}

\begin{figure}[H]
\centering
\includegraphics[width=0.7\linewidth , height=0.5\linewidth]{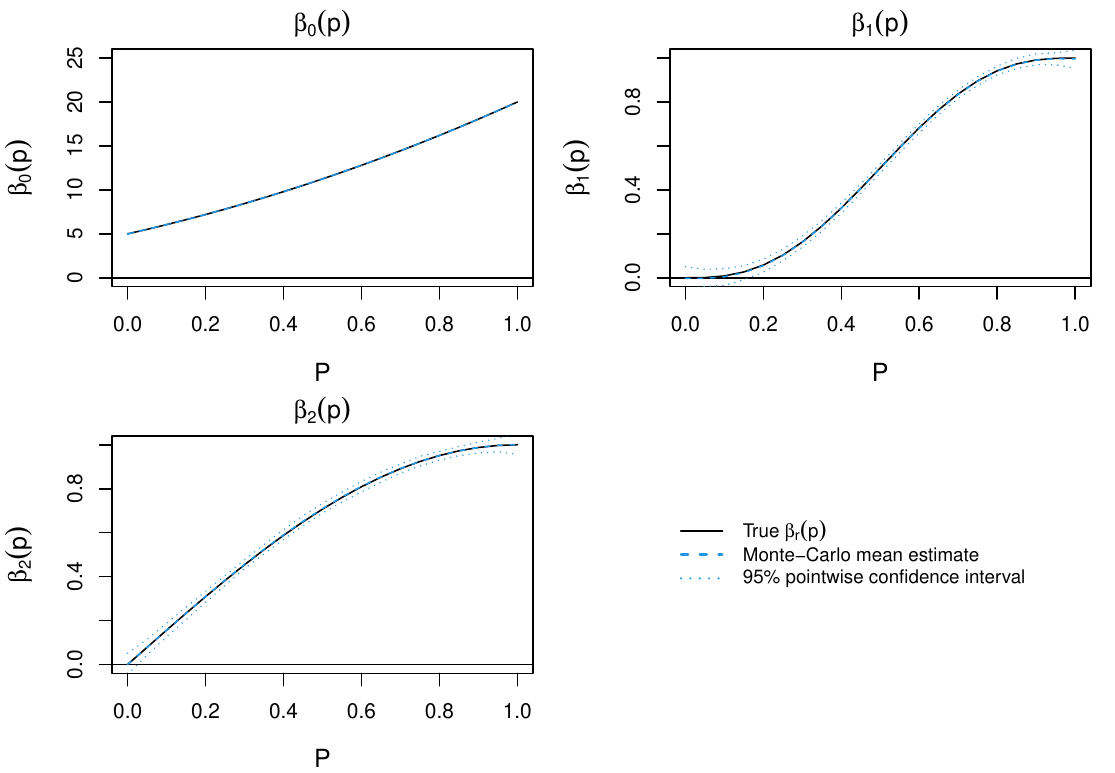}
\caption{The true distributional effects (solid black) and estimated distributional fixed effects (dashed blue) along with its 95$\%$ point-wise confidence intervals (dotted blue), $n=25$.}
\label{fig:fig2}
\end{figure}

We also display the Monte-Carlo mean of the distributional fixed effect estimates, overlaid on the true curve, along with the Monte-Carlo percentile-based confidence intervals in Figure \ref{fig:fig2} for $n=25$. It can be seen that the true distributional fixed effects are estimated with high accuracy. Similar figures for $n=49,100$ are displayed as Figures S2, S3 in the supplementary material. As a measure of predictive accuracy, we also obtain the in-sample mean-squared prediction error, which is equal to the average (across the samples) squared Wasserstein distance in this case. The distribution of this metric across the 100 Monte-Carlo replications is presented in supplementary Figure S4, across the sample sizes and for both SILDOR and DOR. The proposed SILDOR approach can be seen to produce considerably lower squared Wasserstein distance (MSPE) compared to DOR, due to carefully accounting for the spatial and the longitudinal structure.

%computed
%pointwise on $\mathcal{P}$.

\section{Real Data Application}  
\label{sec:real app}
%Data source \url{https://www.epa.gov/castnet/download-data}.

We apply the proposed SILDOR framework to characterize monthly, site-specific
distributions of $\tno$ concentrations across the contiguous United States using data from the Clean Air Status and Trends
Network (CASTNet). Data are publicly available for download at
\url{https://www.epa.gov/castnet/download-data}.
CASTNet measures atmospheric nitrate using a three-stage filter pack that is
exchanged weekly at every monitoring station, yielding a continuous record of weekly $\tno$ (nitric acid plus particulate nitrate)
concentrations. We obtained weekly $\tno$ measurements for the ten-year period
2010-2019 from $n = 92$ CASTNet stations located across the contiguous United
States. The names of the stations along their geographic coordinates are reported in the Supplementary Table S1.  We construct the site and month-specific distributional outcome $\tilde{Q}_{ij}(p)$ by pooling the weekly $\tno$ observations recorded at site $i$ during calendar
month $j$ across all ten years of the study period. These distributional representations capture the long-term monthly distribution of the $\tno$ concentrations at each site \citep{ghosal2025distributional}. The distributional outcome $\tilde{Q}_{ij}(p)$ is evaluated on the equispaced grid $\mathcal{P} = \{0, 0.05, 0.10, \ldots, 1\}$ of length
$S = 21$. The resulting data set comprises  $1{,}094$ site-month-specific distributional
observations across the $92$ stations, with $91-92$ observations recorded
in each of the twelve calendar months. The modest imbalance across sites
reflects station-specific gaps in the ten-year monitoring record and is
accommodated directly by the SILDOR framework, which does not require a
balanced longitudinal design.   

For each site-month, we obtain two meteorological covariates from CASTNet's
co-located hourly meteorological measurements, namely, the monthly average ambient
temperature and the monthly total precipitation.
Both predictors are pooled across the same ten-year window and calendar month as the corresponding $\tno$ distributional outcome. Monthly average temperature is calculated as the mean of all hourly temperature
readings recorded at site $i$ during calendar month $j$ across the ten study years. Monthly total precipitation is calculated as the within-month hourly total averaged across the available years, representing a typical (climatological) monthly total precipitation value at that site.
%by first summing hourly precipitation within each calendar month for a year, and then averaging this monthly total across the years for which data are available at that site.  The resulting monthly total precipitation represents a typical (climatological) monthly precipitation at a particular site.
Each meteorological covariate is then
rescaled to $[0,1]$ via a monotone linear transformation,
\begin{equation*}
  z_{ij1} = \frac{\text{temp}_{ij} - \min_{i,j}\, \text{temp}_{ij}}
                 {\max_{i,j}\, \text{temp}_{ij} - \min_{i,j}\, \text{temp}_{ij}},
  \qquad
  z_{ij2} = \frac{\text{prec}_{ij} - \min_{i,j}\, \text{prec}_{ij}}
                 {\max_{i,j}\, \text{prec}_{ij} - \min_{i,j}\, \text{prec}_{ij}},
\end{equation*}
so that $\mathbf{z}_{ij} = (z_{ij1}, z_{ij2})^{\!\top} \in [0,1]^2$ without loss
of generality, as required in SILDOR.

\begin{figure}[ht]
\centering
\includegraphics[width=0.8\linewidth , height=0.57\linewidth]{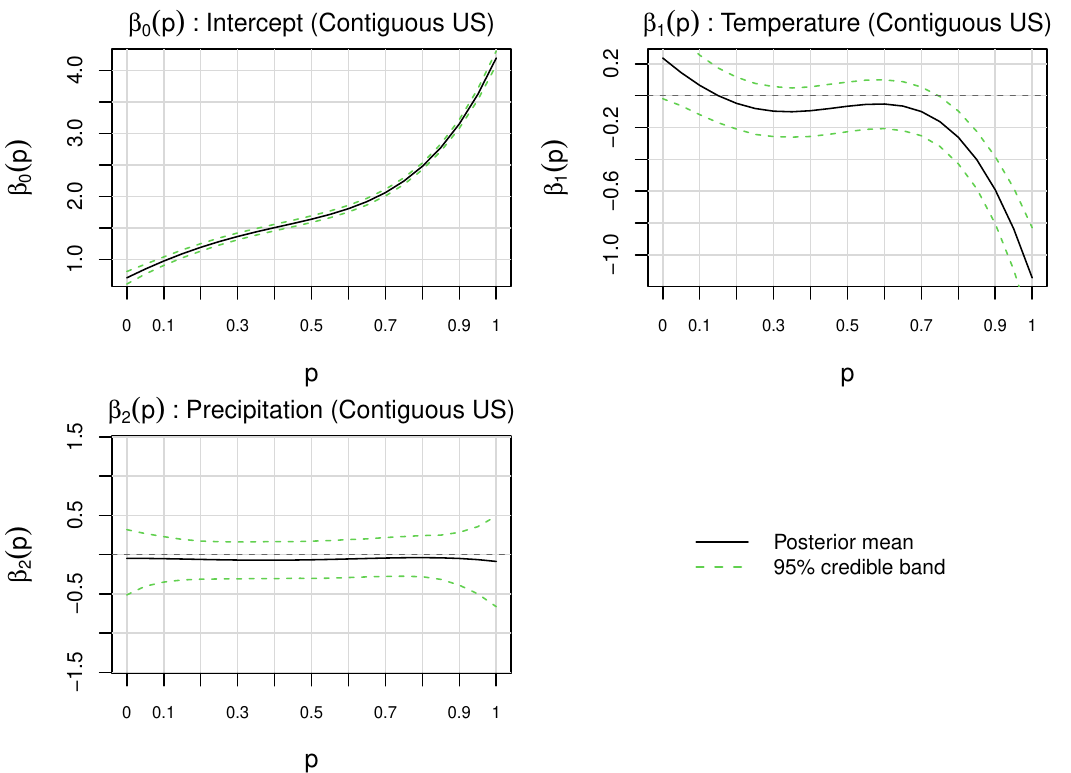}
\caption{Estimated distributional fixed effects (solid) corresponding to the intercept, temperature, and total precipitation along with their 95$\%$ pointwise credible intervals (green dashed) from the proposed SILDOR method. }
\label{fig:figr1}
\end{figure}

The final analytic data set consists of $\tilde{\mathbf{Q}}_{ij}(\cdot)$ observed on the equispaced grid $\mathcal{P}$, together with the covariate pair $\mathbf{z}_{ij} \in
[0,1]^2$, for $n = 92$ sites and up to $J = 12$ calendar months per site,
yielding $1{,}094$ site-month distributional observations in total. We fit the SILDOR model~\eqref{eq:model} to the $\tno$ data described above,
with $q = 2$ scalar covariates corresponding to the rescaled monthly average temperature
($z_{ij1}$) and monthly total precipitation ($z_{ij2}$),
\begin{equation}
  Q_{ij}(p) \;=\; \beta_0(p) \;+\; z_{ij1}\,\beta_1(p) \;+\; z_{ij2}\,\beta_2(p)
              \;+\; W_{ij}(p) \;+\; \epsilon_{ij}(p),
  \qquad p \in [0,1],
  \label{eq:model_app}
\end{equation}
for $i = 1, \ldots, 92$ sites and $j = 1, \ldots, 12$ calendar months. Here
$\beta_0(p)$ is the distributional intercept, $\beta_1(p)$ and $\beta_2(p)$ are
the distributional effects of temperature and precipitation, respectively, on
the $\tno$ distribution at quantile level $p$. We denote by $W_{ij}(p)$ the
spatio-temporal random effect that captures site-and month-specific
deviations from the global mean structure.
The distributional fixed effects $\beta_0(p), \beta_1(p), \beta_2(p)$ are
modeled using a common degree $N=5$ Bernstein polynomial basis
\eqref{eq:bernstein_FE}. For the spatio-temporal random effect $W_{ij}(p)$, we
use $M=3$ Beta-density (Bernstein) basis functions $\psi_m(p)$ in the quantile
direction and $D=7$ cubic B-spline basis functions in the time direction. 
%The orders
%$N$, $M$, and $D$ were selected via the Deviance Information Criterion (DIC) proposed in Section 3.5, comparing candidate models across a grid of basis dimensions and retaining the
%combination that minimizes the DIC.
For the MCMC within SILDOR, we use the diffuse priors specified in Section~\ref{choice}. The
unconstrained Metropolis-within-Gibbs sampler for the MCMC is
run for $N_{\mathrm{iter}} = 40{,}000$ iterations, with a burn-in of
$N_{\mathrm{burn}} = 20{,}000$ and thinning of $N_{\mathrm{thin}} = 5$,
retaining $B = 4{,}000$ posterior draws.  Convergence of the unconstrained MCMC
sampler is assessed via trace plots. Representative trace plots are provided in the
Supplementary Figure S5, illustrating successful convergence. The final fixed effect estimates and the projected-posterior mixed effect prediction of the distributional outcomes are obtained using the projected posterior approaches as illustrated in Section 3.2 and 3.3 of the SILDOR framework.  Figure \ref{fig:figr1} displays the estimated fixed effects $\widehat{\beta}_0(p)$, $\widehat{\beta}_1(p)$, and
$\widehat{\beta}_2(p)$ (projected-posterior mean), together with their pointwise $95\%$ credible intervals from the SILDOR.

%\newpage

%\url{https://pubmed.ncbi.nlm.nih.gov/23682776/}

The estimated intercept $\hat{\beta}_0(p)$ is strictly increasing and convex across all quantile levels, ranging from approximately $0.7\,\mu$g\,m$^{-3}$ at $p = 0$ to $4.0\,\mu$g\,m$^{-3}$ at $p = 1$. The narrow credible band illustrates that the baseline $\tno$ distribution is estimated with high precision. The convex shape of $\hat{\beta}_0(p)$ highlights the positive skewness of $\tno$ distribution across US monitoring stations. The  $\tno$ distribution has a compact lower bulk, and a substantially heavier upper tail, which are driven by episodic high-concentration events concentrated in winter and early spring \citep{ghosh2010spatio, malm2004spatial}.
The estimated temperature effect $\hat{\beta}_1(p)$ is non-monotone across the quantile domain and captures a significant effect of the temperature on the right tail of $\tno$ distribution. The
estimated $\hat{\beta}_1(p)$ is significantly negative at upper quantile levels ($p \geq 0.75$), with the $95\%$ credible band excluding zero for $p$ in the upper tail. The estimated effect is not significant near the central region $p\approx 0.5$, highlighting the advantages of distributional regression over the mean or median regression, where this effect on the upper tail would not be captured. Interestingly, the estimated effect $\hat{\beta}_1(p)$  is positive at lower quantiles, illustrating the varying effect of temperature at different parts of the $\tno$ distribution,  although it is not captured to be significant by the credible bands. This pattern is physically interpretable through the thermodynamics of
gas-particle partitioning. Atmospheric particulate nitrate primarily exists  as
ammonium nitrate (NH$_4$NO$_3$), which is in reversible equilibrium with
gaseous nitric acid (HNO$_3$) and ammonia (NH$_3$):
\begin{equation*}
  \mathrm{NH_4NO_3(s)} \;\rightleftharpoons\; \mathrm{HNO_3(g)} + \mathrm{NH_3(g)}.
\end{equation*}
The equilibrium constant for this reaction is known to be an increasing function
of temperature \citep{stelson1982relative, seinfeld2016atmospheric}.  Elevated
temperatures shift the equilibrium toward the gas phase, volatilizing
particulate NH$_4$NO$_3$ into HNO$_3$ and NH$_3$ and thereby suppressing
particulate nitrate concentrations. This mechanism is particularly active for high $\tno$ concentrations, a favourable condition for this partitioning. Our finding that
$\hat{\beta}_1(p)$ is strongly negative at high quantile levels is therefore
consistent with the well-established principle that formation of particulate
ammonium nitrate is favored under low-temperature conditions \citep{stelson1982relative, seinfeld2016atmospheric}. This finding also matches laboratory evidence that lower temperatures promote particulate
nitrate formation via heterogeneous reactions of NO$_2$ on mineral aerosol
surfaces \citep{wu2013heterogeneous}. Together, these mechanisms explain
why a warmer climate suppresses the extreme upper tail of the $\tno$ distribution while leaving the lower bulk relatively unchanged.

The estimated precipitation effect is marginally negative (close to zero), with the $95\%$ credible band including zero throughout the entire quantile domain. This indicates that total monthly precipitation does not have a significant association with $\tno$ concentrations at any quantile level, after accounting for temperature and the spatio-temporal random effects.  While wet deposition is a known pathway for removal of atmospheric
nitrate \citep{bondell2010joint,ghosh2010spatio}, several factors may explain the null result in
this analysis. Our analysis spans all 92 CASTNet monitoring stations across the contiguous United States, covering
climatologically diverse regions from the humid eastern states to the
arid interior west. In the eastern states near the sea, precipitation is more frequent and chemically active in removing nitrate. However, in the west and mid-west, wet removal is infrequent, and nitrate dynamics are dominated by dry deposition and gas-particle partitioning. 
The opposing
regional signals in the precipitation-nitrate relationship, when pooled into a
single global fixed-effect coefficient, may attenuate each other and yield a
null signal at the national scale. In contrast, earlier analyses focused on
approximately 15 eastern US CASTNet stations \citep{bondell2010joint}, and were able
to detect a more coherent precipitation signal within a climatologically
homogeneous subregion. This suggests that the precipitation-$\tno$
relationship may be spatially non-stationary across the national network. 
To investigate this phenomenon, we performed an additional subanalysis restricting the focus of SILDOR analysis on a subset of  15 eastern US CASTNet stations (See Supplementary Table S2) as in \cite{bondell2010joint}. The distributional effect of precipitation was found to be significant and negative at quantiles $p\in[0.5,0.9]$, highlighting a significant role of wet removal in decreasing $\tno$ concentration in this region (Supplementary Figure S6). We display the SILDOR-predicted (in-sample) monthly site-specific median and 95th percentile of $\tno$ concentrations across the 92 contiguous US CASTNet stations in supplementary Figures S7 and S8. In both cases, SILDOR closely
captures the observed spatio-temporal structure. The predicted surfaces exhibit mild spatial smoothing relative to
the observed data, consistent with the random effect borrowing strength
across neighboring sites. SILDOR can be seen to capture the dominant spatial and seasonal signal well even for extreme quantiles, where the observed data are generally noisiest.

%a spatially varying coefficient extension of SILDOR---in which $\beta_2(p)$
%varies smoothly over geographic space---could recover regional precipitation
%effects masked in the global estimate. We leave this as a direction for future
%work.

%seaboard---where
%precipitation is frequent and chemically active in removing nitrate---
% First, the covariate used here is total monthly precipitation,
%a coarse temporal aggregate that conflates frequent light rainfall events
%(which provide gradual scavenging) and infrequent heavy events (which drive
%episodic washout); the opposing effects of these precipitation regimes may
%cancel at the monthly scale. Second, 

We also compared the estimation and prediction performance of SILDOR against the non-spatial distributional outcome regression (DOR)  \citep{ghosal2025distributional},
which fits the same fixed-effect structure~\eqref{eq:model_app}, but omits the spatio-temporal random effect $W_{ij}(p)$. The fixed-effect estimates from DOR are shown in Supplementary Figure S9. %SILDOR produces narrower $95\%$ credible intervals than the $95\%$ confidence interval produced by DOR (Table \ref{tab:app_bandwidth}).

%\subsection*{Comparsion with DOR estimation, %confidence band width showing uncertainty %reduction}

\begin{table}[H]
\centering
\caption{Average width of pointwise $95\%$ credible/confidence bands for the
estimated distributional coefficients, SILDOR versus the non-spatial DOR
benchmark, CASTNet $\tno$ application.}
\label{tab:app_bandwidth}
\begin{tabular}{lcc}
\toprule
Coefficient & SILDOR & DOR \\
\midrule
$\beta_0(p)$ (Intercept)     &0.126 &0.301 \\
$\beta_1(p)$ (Temperature)   &0.364 &0.501 \\
$\beta_2(p)$ (Precipitation) &0.578 &3.249 \\
\bottomrule
\end{tabular}
\end{table}

SILDOR can be seen (Table \ref{tab:app_bandwidth}).
to yield substantially narrower credible bands than the non-spatial
DOR benchmark for all three distributional coefficients, corresponding to a
$58.7\%$ reduction in average width of the credible band for the intercept $\beta_0(p)$, a $27.5\%$ reduction for the temperature effect $\beta_1(p)$, and a 
$82.5\%$ reduction for the precipitation effect $\beta_2(p)$. By accounting for spatio-temporal dependence, SILDOR provides reduced uncertainty for the estimates of the distributional fixed effects.

%\subsection*{Comparsion of out-of-sample prediction with with DOR}
We compared the out-of-sample predictive accuracy of SILDOR relative to the non-spatial
DOR benchmark at unmonitored spatial locations. We randomly partition the
$92$ stations into a training set comprising $80\%$ of the sites ($74$
stations) and a held-out test set comprising the remaining $20\%$ ($18$
stations), and fit both SILDOR and DOR using only the training-site data. For
SILDOR, the spatio-temporal random effects at the held-out test sites are
drawn from their Gaussian conditional predictive distribution given the
training-site random-effect draws and the estimated Mat\'ern covariance
parameters, following Section 3.4. The projected-posterior mixed-effect prediction framework is then applied at the test sites to obtain valid predicted quantile functions. For DOR, predictions at the held-out sites are obtained from the fitted fixed-effect structure alone, evaluated at the
test-site covariate values, since DOR does not incorporate a spatial
random-effect component. %To avoid information leakage, the covariate
%normalization bounds $[\min, \max]$ used to construct $\mathbf{z}_{ij}$ are
%computed from the training data only and applied unchanged to the test-site
%covariates. 
Predictive accuracy at the held-out stations is summarized using the mean squared
prediction error (MSPE) between the observed and predicted distributional
outcomes at the test sites,
$\mathrm{MSPE} = (n_t J)^{-1} \sum_{i \in \mathrm{test}} \sum_j S^{-1}
\sum_s \{\tilde{Q}_{ij}(p_s) - \widehat{Q}_{ij}(p_s)\}^2$, where $n_t = 18$ is
the number of held-out test sites. This metric approximates the average squared
$2$-Wasserstein distance between the observed and predicted quantile functions
at unmonitored locations. The MSPE for SILDOR is found to be 0.54 compared to a value of 1.24 for DOR, highlighting a $56\%$ reduction in prediction error at the held-out test sites. Supplementary Figure S10 displays the distribution of the
squared Wasserstein distance $S^{-1}\sum_s \{\tilde{Q}_{ij}(p_s) -
\widehat{Q}_{ij}(p_s)\}^2$ between the observed and predicted quantile
functions, computed separately for each held-out site-month combination. The resulting boxplots reflect variability in predictive
accuracy across both space (test sites) and time (calendar months)
simultaneously. SILDOR can be seen to produce a broadly consistent reduction in prediction
error across the held-out stations and months. The interquartile range (IQR) for SILDOR is roughly half that of
DOR, indicating that SILDOR's predictive accuracy is not only better on
average but also considerably more consistent across space and time. DOR,
in contrast, exhibits a much longer upper tail, reflecting a subset of
held-out site-months at which the non-spatial model's prediction performs poorly.
%\subsection*{Representative plots of out-of-sample predictions, showing better prediction with SILDOR}
\begin{figure}[H]
\centering
\includegraphics[width=0.7\linewidth,, height=0.56\linewidth]{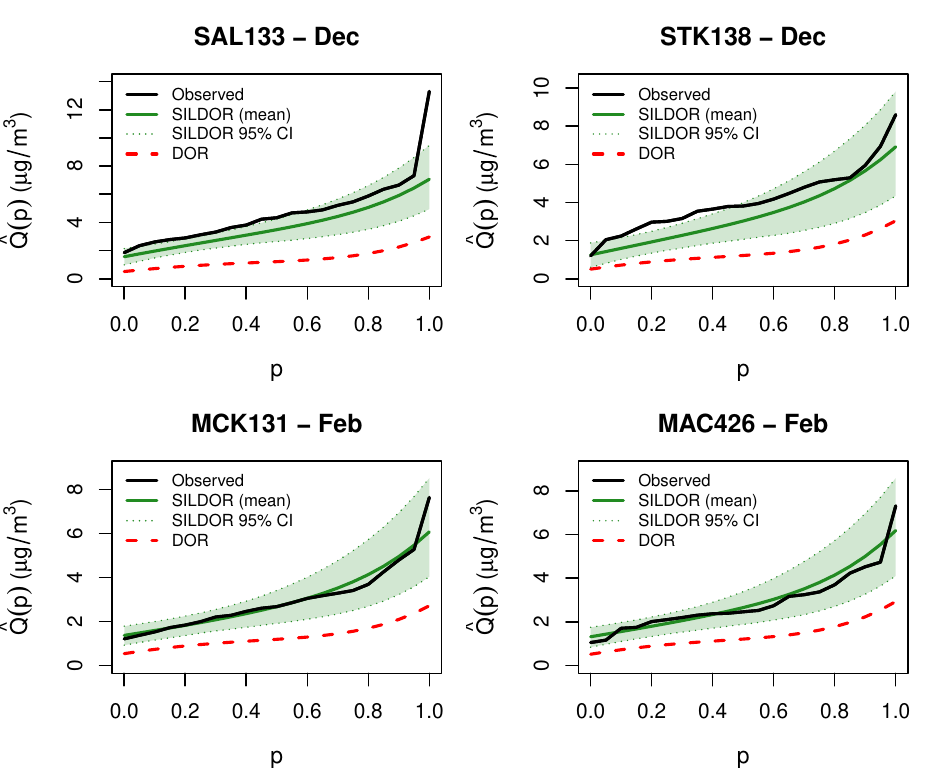}
\caption{Representative out-of-sample predicted quantile functions
$\widehat{Q}(p)$ at four held-out site--month combinations, showing the observed quantile function (solid black), the SILDOR mixed-effect prediction (solid green), and
the non-spatial DOR prediction (dashed red). For SILDOR, the $95\%$ point-wise credible bands are provided for the mixed-effect prediction.}
\label{fig:pred_examples}
\end{figure}

Figure~\ref{fig:pred_examples} displays representative predicted quantile
functions at four held-out site-month combinations, along with their $95\%$ point-wise credible bands.
Across all four examples, the SILDOR-predicted quantile function tracks the
observed quantile function more closely than DOR. DOR consistently
under-predicts $\tno$ concentrations across the entire quantile range and,
in particular, substantially underestimates the upper tail of the
distribution. This closer tracking of the full distributional shape across unmonitored locations underscores the practical benefit of SILDOR. Overall, SILDOR provides markedly better estimation and predictive
performance than the non-spatial DOR benchmark in the CASTNet $\tno$
application. These results underscore the practical value of
explicitly modeling spatial dependence when characterizing and predicting
distributional environmental exposures.

\section{Discussion} \label{sec:discussion}

In this article, we have developed a spatially-indexed longitudinal
distributional outcome regression (SILDOR) model. The distributional fixed effects are modeled using Bernstein polynomial basis
expansions, yielding  interpretable, jointly monotone covariate effects
without requiring monotonicity of individual regression coefficients.
Spatio-temporal random effects are captured through a tensor-product basis expansion
of quantile- and time-direction basis functions allowing flexible modeling of spatio-temporal dependence. A key computational contribution is the efficient two-stage
projected-posterior algorithm, which projects posterior draws onto the constrained
space, guaranteeing valid (non-decreasing) mixed-effect predictions of the
site- and time-specific quantile functions while avoiding the substantial
computational burden of a directly constrained MCMC sampler.

Numerical simulations demonstrated satisfactory estimation and inference performance of SILDOR compared to the non-spatial distributional
outcome regression (DOR) benchmark. We applied SILDOR to characterize the dependence of monthly, station-specific distribution
of total nitrate ($\tno$) concentrations across $92$ CASTNet monitoring
stations on average temperature and total precipitation in the contiguous United States. The estimated temperature effect was
found to be non-monotone across the quantile domain, with a pronounced and
statistically significant suppression of the upper tail of the $\tno$
distribution at higher temperatures.  This finding is  consistent with the
thermodynamics of ammonium nitrate gas-particle partitioning and with
laboratory evidence on the temperature dependence of heterogeneous nitrate
formation.  Relative to the non-spatial DOR benchmark, SILDOR yielded
substantially narrower credible bands for all three distributional
coefficients and more accurate, consistent out-of-sample predictions at
held-out monitoring stations, highlighting the practical value of explicitly
modeling spatial dependence for both inference and prediction in
environmental monitoring applications.

There are certain limitations to the proposed approach that
suggest natural directions for future work. First, SILDOR assumes a linear
(additive) effect of the scalar covariates on the distributional outcome. This may be restrictive in the real world, as the
relationship between a covariate (e.g., temperature) and the distributional outcome may be substantially nonlinear. Moreover, several covariates can have significant interactions among themselves. Extending SILDOR to accommodate nonlinear covariate effects, for
instance through a generalized additive framework or a single-index type model
while preserving the joint monotonicity guarantee, is an important direction
for future methodological development. Second, the distributional fixed effects $\beta_l(p)$, 
are assumed to be fixed (not varying) across all spatial locations, with spatial
heterogeneity entering the model exclusively through the random-effect process
$W_{ij}(p)$. A natural extension of SILDOR would allow the distributional
coefficients themselves to vary smoothly over space. For example, this could be achieved via a
spatially varying coefficient specification of $\beta_l(p; \mathbf{s}_i)$, which can be modeled
with a tensor basis expansion in $\mathbf{s}_i$ and $p$. Such an extension would
allow the model to recover regionally heterogeneous covariate effects while
retaining the interpretability and shape-constrained structure of the current
framework, though at increased computational cost. Finally, the
current application considers only scalar meteorological covariates. Many
environmental monitoring applications also involve distributional predictors, such as 
a co-located distributional pollutant measurements. Extending SILDOR to
accommodate such distributional predictors, in the spirit of the distribution-on-distribution regression component of \citet{ghosal2025distributional}, would further broaden the applicability of the proposed framework to a wider class of spatially-indexed
distributional data problems arising in environmental and public health monitoring, and remains an area of future research.

%%  imsart-nameyear.bst  will be used to                   %%
%%  Use \cite{...} to cite references in text.             %%
%%                                                         %%
%%%%%%%%%%%%%%%%%%%%%%%%%%%%%%%%%%%%%%%%%%%%%%%%%%%%%%%%%%%%%

%% if your bibliography is in bibtex format, uncomment commands:

%\newpage
\bibliographystyle{imsart-nameyear} % Style BST file
\bibliography{refs}       % Bibliography file (usually '*.bib')

\end{document}

\begin{frontmatter}
\title{Supplementary Material for Spatially-Indexed Longitudinal Distributional Outcome Regression for Environmental Monitoring}
\runtitle{Spatially-Indexed Longitudinal DOR}

\begin{aug}
%%%%%%%%%%%%%%%%%%%%%%%%%%%%%%%%%%%%%%%%%%%%%%%
%% Only one address is permitted per author. %%
%% Only division, organization and e-mail is %%
%% included in the address.                  %%
%%%%%%%%%%%%%%%%%%%%%%%%%%%%%%%%%%%%%%%%%%%%%%%

\author[A]{\fnms{Rahul}~\snm{Ghosal}\thanksref{t1}\ead[label=e1]{rghosal@mailbox.sc.edu}},
\author[B]{\fnms{Suman}~\snm{Majumder}\thanksref{t1}\ead[label=e2]{smajumder@isical.ac.in}}
%\and
\author[C]{\fnms{Indranil}~\snm{Sahoo}\ead[label=e3]{sahooi@vcu.edu}}

\thankstext{t1}{Co-first authors.}

%%%%%%%%%%%%%%%%%%%%%%%%%%%%%%%%%%%%%%%%%%%%%%
%% Addresses                                %%
%%%%%%%%%%%%%%%%%%%%%%%%%%%%%%%%%%%%%%%%%%%%%%
\address[A]{Department of Epidemiology and Biostatistics, University of South Carolina, Columbia, SC, USA\printead[presep={,\ }]{e1}}

\address[B]{Interdisciplinary Statistical Research Unit, Indian Statistical Institute, Kolkata, WB, India\printead[presep={,\ }]{e2}}

\address[C]{Department of Statistical Sciences $\&$ Operations Research, Virginia Commonwealth University, Richmond, VA, USA\printead[presep={,\ }]{e3}}
\end{aug}

\begin{keyword}
\kwd{Spatial Distributional Data Analysis}
\kwd{Longitudinal Distributional Outcome Regression}
\kwd{Constrained Optimization}
\kwd{Bayesian}
\kwd{Projected Posterior}
\end{keyword}

\end{frontmatter}

%%%%%%%%%%%%%%%%%%%%%%%%%%%%%%%%%%%%%%%%%%%%%%
%%%% Main text entry area:
%%%%%%%%%%%%%%%%%%%%%%%%%%%%%%%%%%%%%%%%%%%%%%

\section{Appendix A}
\subsection{Proof of Theorem 1}
\begin{theorem}[Monotonicity of the fixed effects prediction]
\label{thm:monotone1}
Define the fixed-effect prediction
$\mu^{\mathrm{F}}(p; \mathbf{z}) = \beta_0(p) + \sum_{l=1}^{q} z_l\,\beta_l(p)$. Then $\mu^{\mathrm{F}}(p; \mathbf{z})$ is non-decreasing in $p$ for every covariate configuration $\mathbf{z} \in [0,1]^q$ if and only if:
\begin{enumerate}[label=\textup{(\roman*)}, noitemsep]
  \item the distributional intercept $\beta_0(p)$ is non-decreasing in $p$.
  \item for every subset $\mathcal{R} \subseteq \{1,\ldots,q\}$, the additive combination $\beta_0(p) + \sum_{l \in \mathcal{R}} \beta_l(p)$ is non-decreasing in $p$.
\end{enumerate}
\end{theorem}
\begin{proof}
 We have $\mu^{\mathrm{F}}(p; \mathbf{z}) = \beta_0(p) + \sum_{l=1}^{q} z_l\,\beta_l(p)$. Since we assume $\beta_l(\cdot)$ are smooth functions, we have 
${\mu^{\mathrm{F}}}^{\prime}(p; \mathbf{z})=\beta_0^{\prime}(p)+\sum_{l=1}^{q}z_l\beta_l^{\prime}(p)$. Enough to show ${\mu^{\mathrm{F}}}^{\prime}(p; \mathbf{z})\geq 0$ for all $(z_1,z_2,\ldots,z_q) \in [0,1]^q$. Note that this is a linear function in $(z_1,z_2,\ldots,z_q) \in [0,1]^q$. By the well-known Bauer's principle, the minimum is attained at the boundary points $B=\{(z_1,z_2,\ldots,z_q):z_j\in \{0,1\}\}$. Hence, the sufficient conditions are given by $\beta_0^{\prime}(p)\geq0$ and  $\beta_0^{\prime}(p)+\sum_{k=1}^{r} \beta_{l_k}^{\prime}(p) \geq0$ for any subset $\{l_1,l_2,\ldots,l_r\}\subset \{1,2,\ldots,q\}$, which follows from condition (i) and (ii). The converse also follows by evaluating $\mu^{\mathrm{F}}(p; \mathbf{z})$ at the vertices of the hypercube.
\end{proof}

\subsection{Proof of Proposition 1}
\begin{proposition}[Monotonicity of the mixed-effect prediction]
\label{prop:mixed}
Let the distributional fixed effects satisfy conditions \textup{(i)-(ii)} of Theorem~\ref{thm:monotone1}. Then the predicted quantile function $\mu_{ij}(p) = \beta_0(p) + \sum_{l=1}^q z_{ijl}\beta_l(p) + W_{ij}(p)$ is non-decreasing in $p$ for every covariate configuration $\mathbf{z} \in [0,1]^q$ if and only if, for every subset $\mathcal{R} \subseteq \{1,\ldots,q\}$,
\begin{equation}
  \beta_0(p) + \sum_{l \in \mathcal{R}} \beta_l(p) + W_{ij}(p)
  \quad\text{is non-decreasing in } p .
  \label{eq:mixed_mono1}
\end{equation}
\end{proposition}

\begin{proof}
For fixed $p$, write $\mu_{ij}(p;\mathbf{z}) = \beta_0(p) + \sum_{l=1}^q
z_l\beta_l(p) + W_{ij}(p)$, which is an affine function of $\mathbf{z} =
(z_1,\ldots,z_q) \in [0,1]^q$ (the term $W_{ij}(p)$ does not depend on
$\mathbf{z}$). Since $[0,1]^q$ is the convex hull of its $2^q$ vertices
$B = \{\mathbf{z} : z_l \in \{0,1\}\}$, any $\mathbf{z} \in [0,1]^q$ can be
written as a convex combination $\mathbf{z} = \sum_{\mathbf{v}\in B}
\lambda_{\mathbf{v}} \mathbf{v}$ with $\lambda_{\mathbf{v}} \geq 0$ and
$\sum_{\mathbf{v}} \lambda_{\mathbf{v}} = 1$. Then by affineness of
$\mu_{ij}(p;\cdot)$ we have,
\begin{equation*}
  \mu_{ij}(p;\mathbf{z}) = \sum_{\mathbf{v} \in B} \lambda_{\mathbf{v}}\,
  \mu_{ij}(p;\mathbf{v}).
\end{equation*}
Each vertex $\mathbf{v} \in B$ corresponds to a subset $\mathcal{R}
\subseteq \{1,\ldots,q\}$ (with $z_l = 1$ iff $l \in \mathcal{R}$), so
$\mu_{ij}(p;\mathbf{v}) = \beta_0(p) + \sum_{l\in\mathcal{R}}\beta_l(p) +
W_{ij}(p)$, which is non-decreasing in $p$ by \eqref{eq:mixed_mono1}. Since
a convex combination (with fixed, nonnegative weights $\lambda_{\mathbf v}$)
of functions non-decreasing in $p$ is itself non-decreasing in $p$, and the
weights $\lambda_{\mathbf v}$ do not depend on $p$, $\mu_{ij}(p;\mathbf{z})$
is non-decreasing in $p$ for every $\mathbf{z} \in [0,1]^q$. Conversely, if
$\mu_{ij}(p;\mathbf{z})$ is non-decreasing in $p$ for every $\mathbf{z} \in
[0,1]^q$, this holds in particular at each vertex $\mathbf{v} \in B$, which
gives exactly \eqref{eq:mixed_mono1} for every subset $\mathcal{R} \subseteq
\{1,\ldots,q\}$.
\end{proof}

\subsection{Formation of the constraint matrix}
We model each distributional coefficient $\beta_l(p)$, $l = 0,1,\ldots,q$,
using a Bernstein polynomial basis of degree $N$,
\begin{equation}
  \beta_l(p) \;=\; \sum_{k=0}^{N} \beta_{lk}\, b_k(p; N),
  \qquad
  b_k(p; N) = \binom{N}{k} p^k (1-p)^{N-k},
  \label{eq:bernstein_FE}
\end{equation}
with basis-coefficient vector $\bm{\beta}_l = (\beta_{l0}, \ldots,
\beta_{lN})^{\!\top}$. The monotonicity of a Bernstein expansion is equivalent to a set
of linear inequality constraints on its coefficients \citep{wang2012shape,ghosal2023shape}.

Specifically, a Bernstein basis expansion
$f_N(p)=\sum_{k=0}^N c_k b_k(p;N)$ is non-decreasing on $[0,1]$ if
$\mathbf{A}_1 \mathbf{c} \ge \mathbf{0}$, where $\mathbf{c} = (c_0,\ldots,
c_N)^{\!\top}$, and $\mathbf{A}_1 \in \mathbb{R}^{N \times (N+1)}$ is the
first-difference matrix with $[\mathbf{A}_1]_{k,k} = -1$ and
$[\mathbf{A}_1]_{k,k+1} = 1$. This can be seen by observing, $f_N^{\prime}(p)=N\sum_{k=0}^{N-1}(c_{k+1}-c_{k})b_k(p,N-1).$ Hence if $c_{k+1}\geq c_{k}$ for $k=0,1,\ldots, N-1$, $f_{N}(x)$ is non decreasing, which is achieved with the constraint matrix $\mathbf{A}_1$.

Stacking the fixed-effect coefficients into
$\bm{\psi} = (\bm{\beta}_0^{\!\top}, \bm{\beta}_1^{\!\top}, \ldots,
\bm{\beta}_q^{\!\top})^{\!\top}$, conditions~(i)--(ii) of
Theorem 1 in the paper translate into the single linear-inequality system
\begin{equation}
  \mathbf{A}\,\bm{\psi} \;\ge\; \mathbf{0},
  \qquad
  \mathbf{A} = \mathbf{T} \otimes \mathbf{A}_1,
  \label{eq:constraint_FE}
\end{equation}
where the rows of the $\{0,1\}$ indicator matrix $\mathbf{T}$ enumerate the
$2^q$ additive combinations of the slopes with the intercept corresponding to Theorem 1. 

For example, when $q = 2$, the
four relevant combinations are $\{\beta_0(\cdot)\}$, $\{\beta_0(\cdot) + \beta_1(\cdot)\}$,
$\{\beta_0(\cdot) + \beta_2(\cdot)\}$, and $\{\beta_0 (\cdot)+ \beta_1(\cdot) + \beta_2(\cdot)\}$, giving
\begin{equation}
  \mathbf{T} =
  \begin{pmatrix}
    1 & 0 & 0 \\
    1 & 1 & 0 \\
    1 & 0 & 1 \\
    1 & 1 & 1
  \end{pmatrix},
  \qquad
  \mathbf{A} =
  \begin{pmatrix}
    \mathbf{A}_1 & \mathbf{0}    & \mathbf{0}    \\
    \mathbf{A}_1 & \mathbf{A}_1  & \mathbf{0}    \\
    \mathbf{A}_1 & \mathbf{0}    & \mathbf{A}_1  \\
    \mathbf{A}_1 & \mathbf{A}_1  & \mathbf{A}_1
  \end{pmatrix}.
  \label{eq:Tmat}
\end{equation}

\section{Appendix B}

\clearpage
\newpage
\FloatBarrier

\section{Supplementary Tables} 

\begin{longtable}{lcc}
\caption{CASTNet station identifiers and geographic coordinates
for the $n = 92$ monitoring stations used in the $\tno$ application.}
\label{tab:stations} \\
\toprule
Site ID & Latitude & Longitude \\
\midrule
\endfirsthead
\toprule
Site ID & Latitude & Longitude \\
\midrule
\endhead
\bottomrule
\endfoot
ABT147 & 41.8405 & -72.0104 \\
ACA416 & 44.3771 & -68.2608 \\
ALB801 & 53.6824 & -112.8680 \\
ALC188 & 30.7016 & -94.6740 \\
ANA115 & 42.4166 & -83.9022 \\
ARE128 & 39.9232 & -77.3079 \\
ASH135 & 46.6038 & -68.4132 \\
BAS601 & 44.2800 & -108.0411 \\
BBE401 & 29.3027 & -103.1778 \\
BEL116 & 39.0282 & -76.8171 \\
BFT142 & 34.8847 & -76.6207 \\
BUF603 & 44.1442 & -106.1089 \\
BVL130 & 40.0520 & -88.3725 \\
BWR139 & 38.4450 & -76.1113 \\
CAD150 & 34.1793 & -93.0988 \\
CAN407 & 38.4583 & -109.8213 \\
CAT175 & 41.9423 & -74.5520 \\
CDR119 & 38.8795 & -80.8477 \\
CDZ171 & 36.7841 & -87.8502 \\
CHA467 & 32.0094 & -109.3891 \\
CHE185 & 35.7508 & -94.6698 \\
CKT136 & 37.9215 & -83.0663 \\
CND125 & 35.2633 & -79.8375 \\
CNT169 & 41.3645 & -106.2400 \\
COW137 & 35.0605 & -83.4303 \\
CTH110 & 42.4009 & -76.6535 \\
CVL151 & 34.0027 & -89.7992 \\
DCP114 & 39.6359 & -83.2606 \\
DIN431 & 40.4373 & -109.3046 \\
DUK008 & 35.9745 & -79.0990 \\
EGB181 & 44.2311 & -79.7831 \\
ESP127 & 36.0389 & -85.7331 \\
EVE419 & 25.3912 & -80.6808 \\
FOR605 & 44.3395 & -105.9198 \\
GAS153 & 33.1812 & -84.4101 \\
GLR468 & 48.5103 & -113.9968 \\
GRB411 & 39.0051 & -114.2159 \\
GRC474 & 36.0586 & -112.1836 \\
GRS420 & 35.6335 & -83.9416 \\
GTH161 & 38.9563 & -106.9859 \\
HOX148 & 44.1809 & -85.7390 \\
HWF187 & 43.9730 & -74.2233 \\
IRL141 & 27.8492 & -80.4556 \\
JOT403 & 34.0696 & -116.3889 \\
KEF112 & 41.5981 & -78.7679 \\
KIC003 & 39.8539 & -95.6578 \\
KNZ184 & 39.1022 & -96.6096 \\
LAV410 & 40.5400 & -121.5765 \\
LRL117 & 39.9883 & -79.2516 \\
MAC426 & 37.1318 & -86.1430 \\
MCK131 & 37.7047 & -85.0487 \\
MCK231 & 37.7047 & -85.0487 \\
MEV405 & 37.1984 & -108.4905 \\
MKG113 & 41.4268 & -80.1452 \\
NEC602 & 43.8730 & -104.1919 \\
NIC001 & 43.6805 & -74.9891 \\
NPT006 & 46.2756 & -116.0216 \\
OXF122 & 39.5311 & -84.7235 \\
PAL190 & 34.8806 & -101.6647 \\
PAR107 & 39.0904 & -79.6617 \\
PED108 & 37.1652 & -78.3071 \\
PET427 & 34.8225 & -109.8925 \\
PIN414 & 36.4832 & -121.1569 \\
PND165 & 42.9290 & -109.7878 \\
PNF126 & 36.1054 & -82.0450 \\
PRK134 & 45.2065 & -90.5972 \\
PSU106 & 40.7209 & -77.9318 \\
QAK172 & 39.9427 & -81.3379 \\
RED004 & 47.8638 & -94.8352 \\
ROM206 & 40.2781 & -105.5456 \\
ROM406 & 40.2781 & -105.5456 \\
SAL133 & 40.8160 & -85.6614 \\
SAN189 & 42.8292 & -97.8541 \\
SEK430 & 36.4895 & -118.8292 \\
SHE604 & 44.9300 & -106.8500 \\
SHN418 & 38.5231 & -78.4347 \\
SND152 & 34.2890 & -85.9701 \\
SPD111 & 36.4698 & -83.8265 \\
STK138 & 42.2872 & -90.0000 \\
SUM156 & 30.1102 & -84.9904 \\
THR422 & 46.8948 & -103.3777 \\
UND002 & 44.5283 & -72.8688 \\
UVL124 & 43.6136 & -83.3599 \\
VIN140 & 38.7408 & -87.4849 \\
VOY413 & 48.4125 & -92.8292 \\
VPI120 & 37.3232 & -80.4572 \\
WFM105 & 44.3900 & -73.8600 \\
WNC429 & 43.5576 & -103.4839 \\
WSP144 & 40.3123 & -74.8727 \\
WST109 & 43.9445 & -71.7008 \\
YEL408 & 44.5654 & -110.4003 \\
YOS404 & 37.7133 & -119.7062 \\
\end{longtable}

\begin{longtable}{lcc}
\caption{CASTNet station identifiers and geographic coordinates
for the $n = 15$ eastern monitoring stations used in the eastern
subset analysis.}
\label{tab:stations_eastern15} \\
\toprule
Site ID & Latitude & Longitude \\
\midrule
\endfirsthead
\toprule
Site ID & Latitude & Longitude \\
\midrule
\endhead
\bottomrule
\endfoot
ANA115 & 42.4166 & -83.9022 \\
BEL116 & 39.0282 & -76.8171 \\
CAD150 & 34.1793 & -93.0988 \\
CDR119 & 38.8795 & -80.8477 \\
CND125 & 35.2633 & -79.8375 \\
COW137 & 35.0605 & -83.4303 \\
CTH110 & 42.4009 & -76.6535 \\
DCP114 & 39.6359 & -83.2606 \\
ESP127 & 36.0389 & -85.7331 \\
GAS153 & 33.1812 & -84.4101 \\
MKG113 & 41.4268 & -80.1452 \\
PNF126 & 36.1054 & -82.0450 \\
PSU106 & 40.7209 & -77.9318 \\
SHN418 & 38.5231 & -78.4347 \\
VPI120 & 37.3232 & -80.4572 \\
\end{longtable}
\clearpage
\newpage
\section{Supplementary Figures}
%\FloatBarrier
%\section{Supplementary Figures}

\begin{figure}[H]
\centering
\includegraphics[width=0.9\linewidth , height=0.62\linewidth]{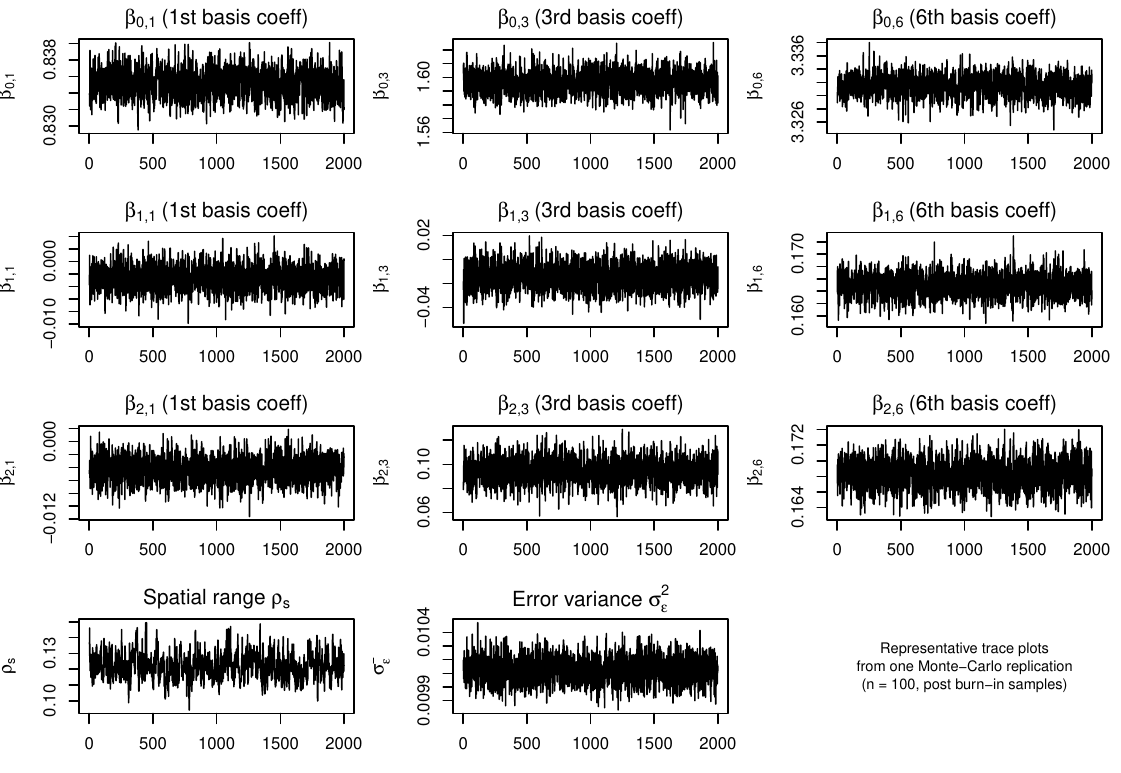}
\caption{Representative trace plots from one Monte-Carlo replication for a few selected parameters, $n=100$ (post burn-in samples).}
\label{fig:fig2}
\end{figure}
\newpage

\begin{figure}[H]
\centering
\includegraphics[width=0.9\linewidth , height=0.62\linewidth]{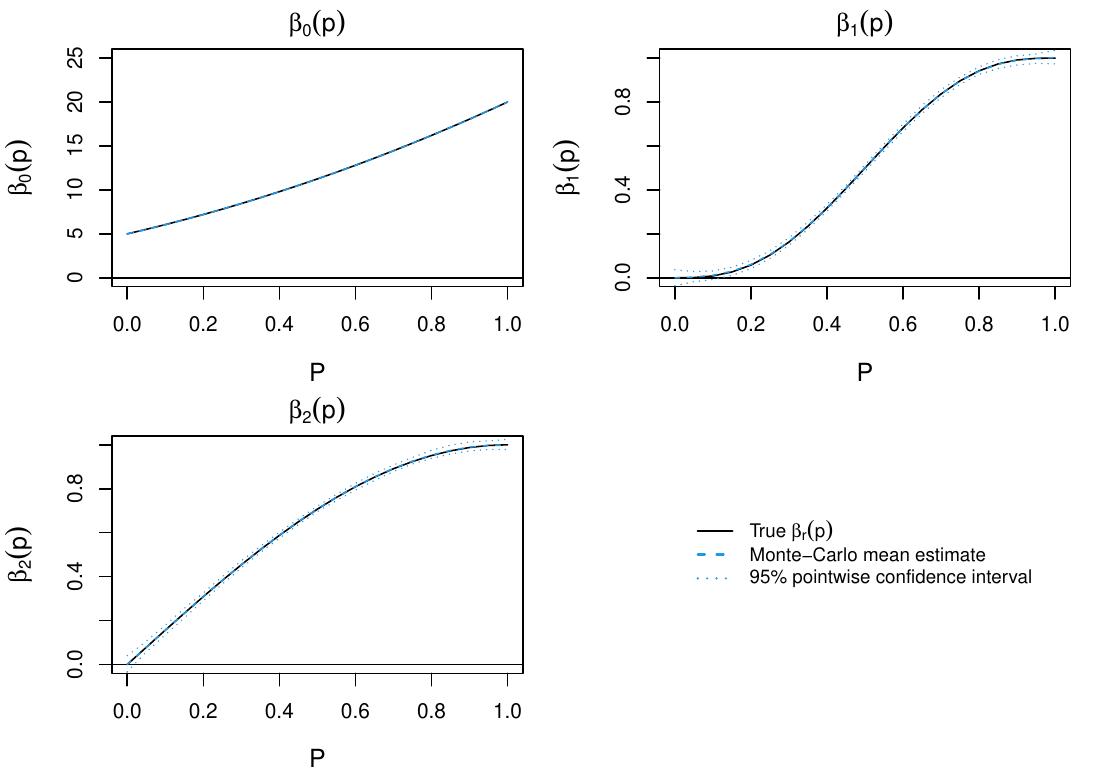}
\caption{Estimated distributional fixed effects (dashed blue) overlaid on the true distributional effects (solid black), along with their 95$\%$ point-wise confidence intervals (dotted blue), $n=49$. }
\label{fig:fig2}
\end{figure}

\begin{figure}[H]
\centering
\includegraphics[width=0.9\linewidth , height=0.62\linewidth]{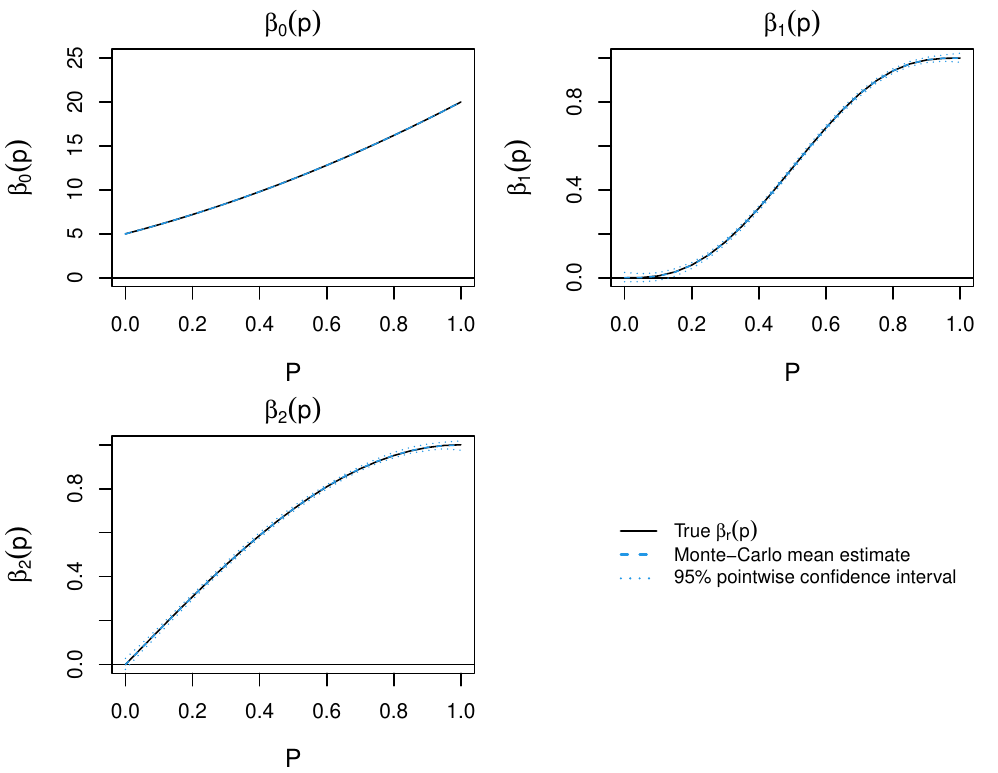}
\caption{Estimated distributional fixed effects (dashed blue) overlaid on the true distributional effects (solid black), along with their 95$\%$ point-wise confidence intervals (dotted blue), $n=100$. }
\label{fig:fig3}
\end{figure}

\begin{figure}[H]
\centering
\includegraphics[width=0.9\linewidth , height=0.62\linewidth]{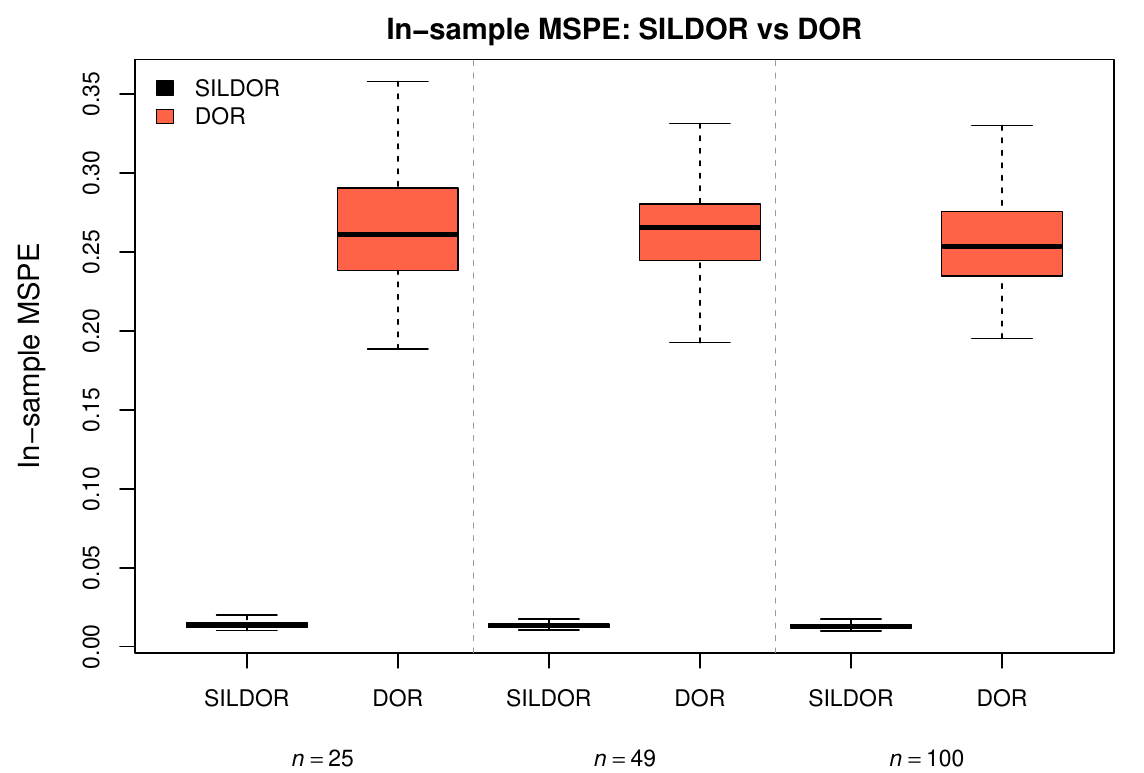}
\caption{Comparison of in-sample MSPE between SILDOR and the DOR method. Displayed are the distributions of in-sample MSPE (average Wasserstein distance) across 100 Monte-Carlo replications for all spatial location sizes $n=25,49,100$. }
\label{fig:fig4}
\end{figure}

\begin{figure}[H]
\centering
\includegraphics[width=0.9\linewidth , height=0.62\linewidth]{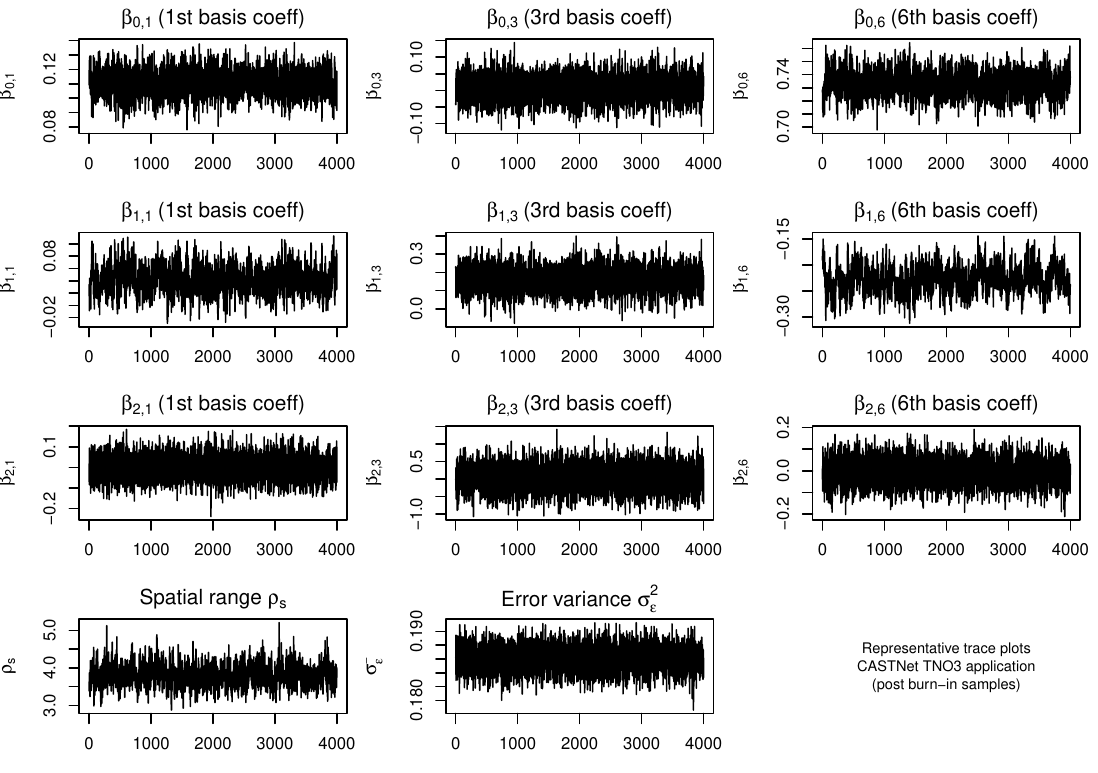}
\caption{Representative trace plots for a few selected parameters in the real data application, post burn-in samples.}
\label{fig:fig2}
\end{figure}
\newpage

\begin{figure}[H]
\centering
\includegraphics[width=0.8\linewidth , height=0.6\linewidth]{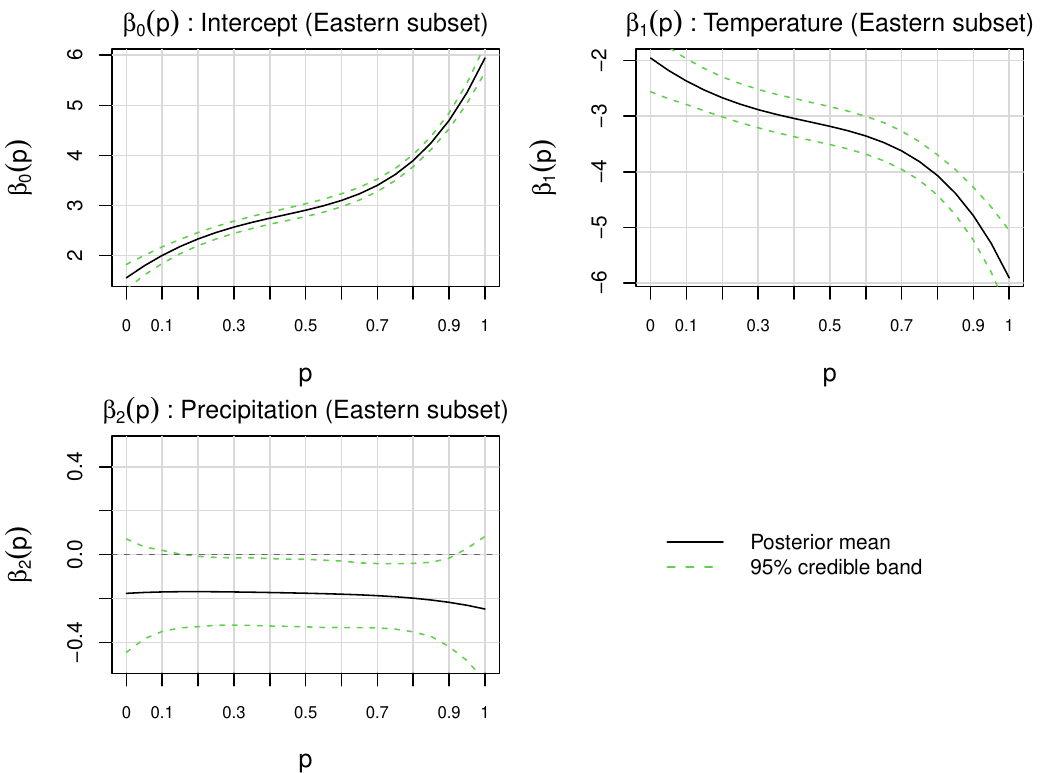}
\caption{Estimated distributional fixed effects (solid) corresponding to the intercept, temperature, and total precipitation along with their 95$\%$ pointwise credible intervals (dotted blue) from the proposed SILDOR method applied to the subset of 15 eastern US CASTNet stations. }
\label{fig:figr1}
\end{figure}

\begin{figure}[H]
\centering
\includegraphics[width=1.1\linewidth , height=0.7\linewidth]{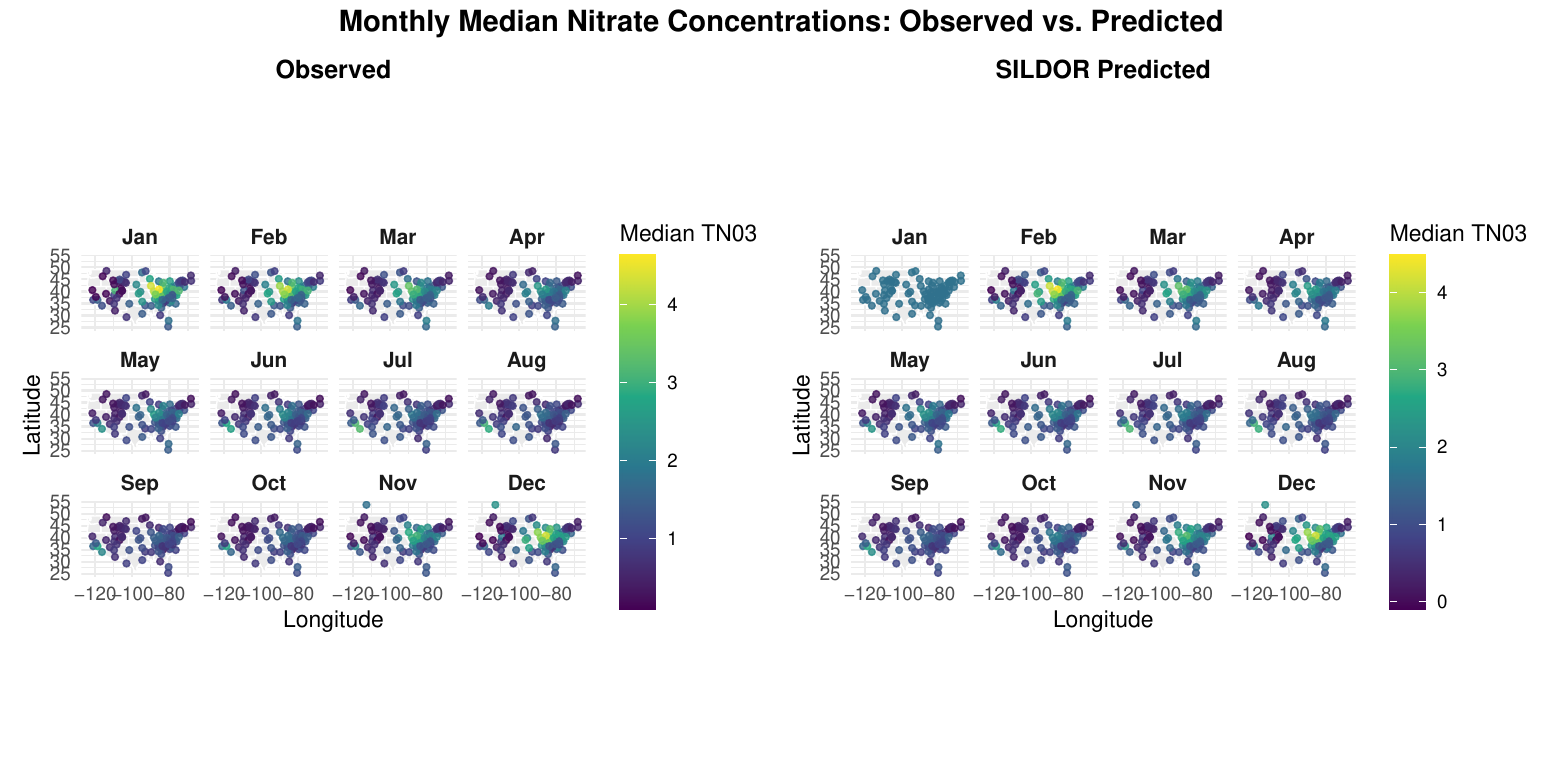}
\caption{Monthly site-specific observed (left) and SILDOR -predicted (in-sample) median $\tno$ concentrations (right)  across the 92 contiguous US
CASTNet stations.}
\label{fig:median_pred}
\end{figure}

\begin{figure}[H]
\centering
\includegraphics[width=1.1\linewidth , height=0.7\linewidth]{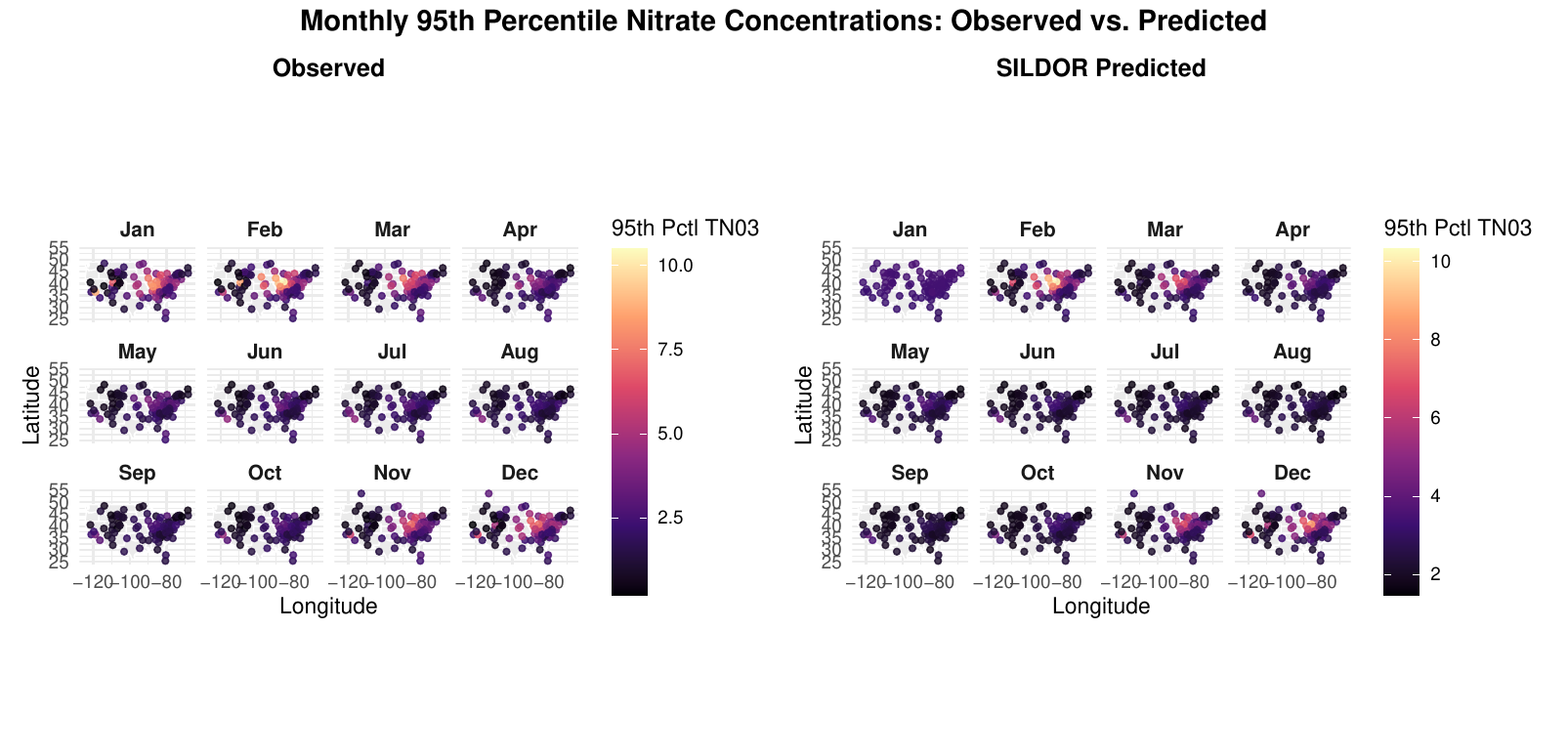}
\caption{Monthly site-specific observed (left) and SILDOR -predicted (in-sample) $95th$ percentile of $\tno$ concentrations (right)  across the 92 contiguous US
CASTNet stations.}
\label{fig:median_pred}
\end{figure}

\begin{figure}[H]
\centering
\includegraphics[width=0.9\linewidth , height=0.8\linewidth]{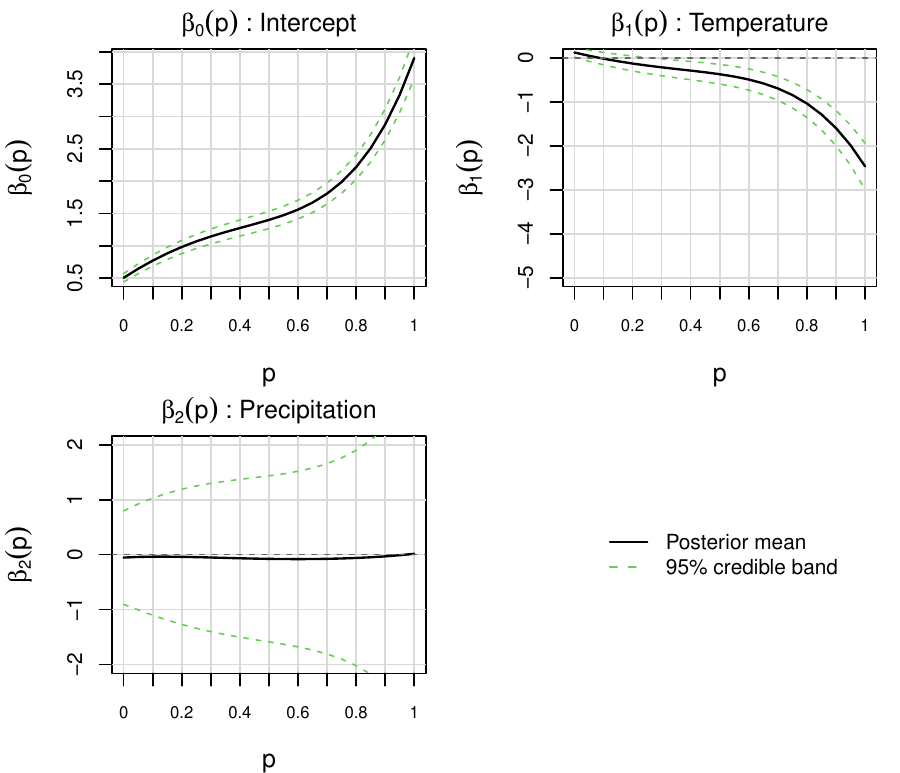}
\caption{Estimated distributional fixed effects (solid) corresponding to the intercept, temperature, and total precipitation along with their 95$\%$ pointwise confidence intervals (dotted blue) from the DOR method (non-spatial). }
\label{fig:figs5}
\end{figure}

\begin{figure}[H]
\centering
\includegraphics[width=0.8\linewidth , height=0.6\linewidth]{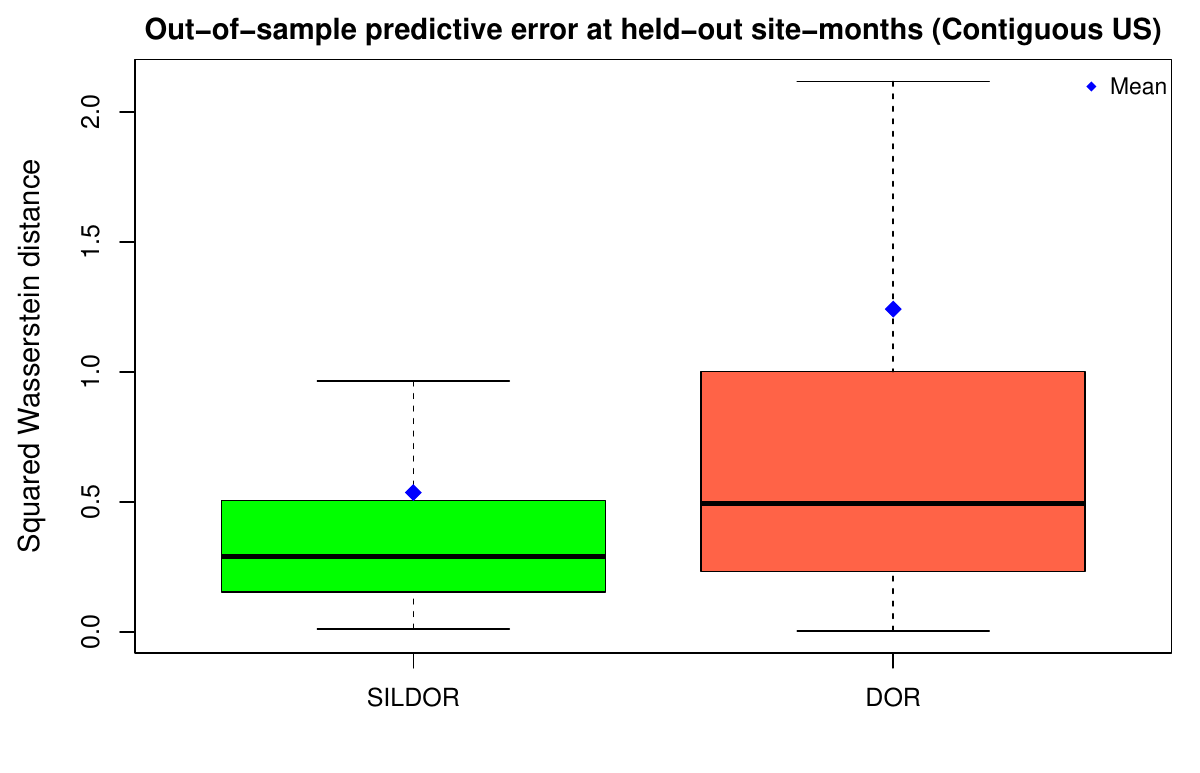}
\caption{Displayed are the distribution of out-of-sample squared Wasserstein distances at held-out site-months from SILDOR and DOR. }
\label{fig:figs6}
\end{figure}

%%  imsart-nameyear.bst  will be used to                   %%
%%  Use \cite{...} to cite references in text.             %%
%%                                                         %%
%%%%%%%%%%%%%%%%%%%%%%%%%%%%%%%%%%%%%%%%%%%%%%%%%%%%%%%%%%%%%

%% if your bibliography is in bibtex format, uncomment commands:

%\newpage
\bibliographystyle{imsart-nameyear} % Style BST file
\bibliography{refs}       % Bibliography file (usually '*.bib')